# High-resolution hierarchical PV system performance modeling in urban environments

B. Tian[*, 1], R.C.G.M. Loonen[1], R.M.E. Valckenborg[2], J.L.M. Hensen[1]

[1] Building Physics and Services, Eindhoven University of Technology, Postbus 513, 5600 MB Eindhoven, the Netherlands

[2] Netherlands Organization for Scientific Research (TNO), High Tech Campus 21, 5656 AE Eindhoven, the Netherlands

## Abstract

Accurate performance modeling of PV systems in urban environments is a significant challenge due to complex partial shading. This study introduces a high-resolution, hierarchical modeling framework that provides detailed insights from the solar cell to the system level. Rigorously validated against field-test data from calibrated equipment, the model demonstrates high accuracy in predicting minute-wised dynamic electrical characteristics ($R^2 > 0.90$). A key finding is the critical shortcoming of conventional, coarser-resolution models under realistic shading; these are shown to overestimate the actual string operating power by up to 163% and the monthly energy yield by up to 54%. The proposed framework avoids these errors by precisely capturing mismatch losses and the time-varying phenomena of system components, such as bypass diode activations. Furthermore, the model accurately quantifies the effectiveness of mitigation technologies, showing that Module-Level Power Electronics (MLPEs) can increase the monthly energy yield of a heavily shaded string by over 20%. This research provides a crucial tool for reliable system design, accurate power forecasting, and the optimization of PV systems in complex urban settings.

## Abbreviation list

| | |
|---|---|
| BIPV | Building Integrated PV |
| CAD | Computer-Aided Design |
| DGCNN | Dynamic Graph Convolutional Neural Network |
| DSO | Distribution System Operator |
| GIS | Geographic Information System |

---

[*] Corresponding author.
  E-mail address: b.tian@tue.nl (B. Tian).

| KNMI | Koninklijk Nederlands Meteorologisch Instituut |
| --- | --- |
| LiDAR | Light Detection and Ranging |
| MBE | Mean Bias Error |
| MLPE | Module-Level Power Electronics |
| MPP | Maximum Power Point |
| MPPT | Maximum Power Point Tracking |
| OPP | Operation Power Point |
| PV | Photovoltaic |
| RMSE | Root Mean Squared Error |

# 1. Introduction

The global shift to renewable energy is accelerating, with solar photovoltaic (PV) technology playing an important role. In 2024, a surge in electricity demand was met almost entirely by the record-breaking expansion of low-emission sources, led by solar PV [1]. This trend is expected to continue, with renewables on track to surpass coal as the largest source of global electricity in 2025 [2].

Within this transition, distributed PV systems in urban areas have become a primary driver of growth [3]. These systems are projected to be a key factor as solar PV becomes the second-largest low-emissions electricity source by 2027 [2], and account for nearly half of the total solar PV expansion through 2030 [4]. The rapid adoption of distributed PV is largely driven by its increasing economic attractiveness for commercial, agricultural, and industrial consumers aiming to lower their energy costs [4,5].

*1.1 PV system performance modeling*

Integrating PV systems into urban landscapes presents unique challenges. One important issue concerns partial shading impacts caused by obstructions such as urban infrastructures and vegetation [6]. Shading not only reduces the potential energy yield but also introduces dynamic power mismatch losses, significantly impairing PV performance [7]. Partial shading is of particular relevance in dense residential neighborhoods, where high energy bills lead to increased interest in adopting distributed PV, but the optimal system placement is challenged by the available roof area and surrounding obstruction factors [8,9]. Conflicts often arise between maximizing household PV system yield and adhering to neighborhood landscape planning, sometimes leading to drastic measures like tree vandalism to reduce shading effects [10]. Therefore, system performance modeling is necessary to aid decision-making during the design and planning phase.

Beyond the individual system level, the rapid growth of decentralized PV generation poses significant challenges for the stability and reliability of the urban power grid. The dynamic and intermittent nature of PV power, which is highly dependent on geographical location and local weather conditions, makes forecasting the total aggregated output difficult [11]. This problem is compounded by the fact that a significant portion of generated electricity is self-consumed within buildings and is therefore not measured by distribution system operators (DSOs). This lack of insight into the true timing and quantity of PV generation creates major challenges for DSOs regarding grid balancing and capacity planning [12]. In this context, solar PV power modeling has also emerged as an essential tool to forecast and estimate generation across wide areas, thereby helping to address these critical grid integration issues.

### 1.1.1 PV modeling frameworks

PV modeling is a complex multi-physics problem that includes the irradiance transfer from sky to PV surface, the conversion between irradiance and electricity, and the interaction between different components in the PV circuit. In practice, the multi-physics problem can be solved by constructing a complete PV modeling framework. A key differentiator in the existing modeling frameworks lies in their spatial modeling resolution, which is crucial for capturing the system performance across diverse environments. Spatial resolution is determined by the smallest system unit included in computations, varying from individual solar cells to entire systems. This resolution affects how well a model can simulate real-world conditions, especially in urban areas where shading patterns and installation orientations greatly influence energy production. For instance, a detailed comparison study by McCarty et al. investigated eight simulation frameworks and found that the choice of spatial resolution is most critical in sparsely shaded conditions, where predicted yields varied significantly between models [8]. The study further revealed that the results from low resolution frameworks and high resolution framework have significant sensitivity variations to the shadow shapes and coverages from the nearby solar obstructions. Given that this choice of spatial resolution is fundamental to a model's accuracy and behavior, mainstream PV modeling frameworks can be broadly categorized into three types: cell-level resolution, module-level resolution, and string-level resolution.

Table 1. Review of cell-level PV modeling frameworks.

| Refs. | Complete framework | Input geometry | Input availability | Irradiance estimation method | Power calculation model | Irradiance resolution | Power resolution | Partial shading definition | Hierarchical System modeling | Support MLPEs |
|---|---|---|---|---|---|---|---|---|---|---|
| [13–16] | ○ | - | - | Pre-defined | One-diode | Module level | Cell level | - | ○ | ○ |
| [17] | ○ | - | - | Pre-defined | Two-diode | Module level | Cell level | - | ○ | ○ |
| [18] | ○ | - | - | Pre-defined | One-diode | Module level | Cell level | Manual setup | ● | ○ |
| [19,20] | ○ | - | - | Pre-defined | One-diode | Cell level | Cell level | Manual setup | ● | ○ |
| [21] | ● | - | - | Transposition | One-diode | String level | Cell level | - | ○ | ○ |
| [22] | ● | CAD | Low | Raytracing | Two-diode | Cell level | Cell level | From geometry | ● | ○ |
| [23] | ● | CAD | Low | Raytracing | One-diode | Cell level | Cell level | From geometry | ● | ○ |
| [24] | ● | CAD | Low | Measurement | One-diode | Cell level | Cell level | Measurement | ○ | ○ |
| [25] | ● | - | - | Measurement | One-diode | Cell level | Cell level | Measurement | ● | ○ |

*Cell-level Resolution Modeling.* Cell-level resolution modeling adopts a detailed approach, using the PV cell as the fundamental unit of calculation. This method models each cell as an equivalent diode circuit, considering unique electrical characteristics of different types of solar cells [13,14]. Representative studies for PV modeling cell-level power outputs are reviewed in Table 1. This high level of detail of cell-level modeling facilitates an examination of micro-level phenomena like partial shading and localized temperature variations, which are crucial for understanding individual cell performance and their collective impact on the overall system [19].

However, a major gap in cell-level modeling studies is the incomplete modeling framework, i.e., the lack of access to system installation information, environmental factors, and irradiance data, which allows the model to focus narrowly on standalone module performance under given irradiance and fails to provide a complete description of the multi-physics problem involved in PV generation [13–17]. Besides, in an incomplete modeling framework, the partial shading impacts can only be defined in a rough way by human interventions [18–20] instead of by analyzing surrounding geometries, which reduces the applicability of the model while greatly increasing the modeling effort required, especially when dealing with dynamic shading situations.

Referring to Table 1, another factor that limits the real-world application of cell-level modeling frameworks is the availability of modeling inputs. Some complete modeling frameworks require the use of high-precision Computer-Aided Design (CAD) models to define the modeled scene [22,23], which could be a labor-intensive task due to the urban environments commonly having high geometrical complexity [26]. Furthermore, some modeling frameworks also require actual irradiance measurements as input [24,25], which limits their applicability in the design phase and in cases lacking reference measurement databases.

Finally, resolution mismatch problems can be identified in some modeling frameworks, where instead of adopting high resolution irradiance models (e.g., raytracing methods [27]) to obtain irradiance distribution at cell-level, low-resolution irradiance data at module-level (module-averaged) [13–18] or string-level (string-averaged) [21] are used as computational inputs to the cell-level power models. This mismatch between the resolution of the irradiance input and the power model has been shown to be a source of uncertainties, being most pronounced when simulating individual module's maximum power outputs [8].

Table 2. Review of module-level PV modeling frameworks.

| Refs. | Complete framework | Input geometry | Input availability | Irradiance estimation method | Power calculation model | Irradiance resolution | Power resolution | Partial shading definition | Hierarchical System modeling | Support MLPEs |
|---|---|---|---|---|---|---|---|---|---|---|
| [28,29] | ○ | - | - | Transposition | PVWatts | Module level | Module level | - | ○ | ○ |
| [30] | ○ | - | - | Transposition | SAPM | Module level | Module level | - | ○ | ○ |
| [31] | ○ | - | - | Measurement | PVWatts | Module level | Module level | - | ○ | ○ |
| [32] | ○ | - | - | Measurement | SAPM | Module level | Module level | - | ○ | ○ |
| [33] | ● | CAD | Low | Transposition | SAPM | Module level | Module level | - | ○ | ○ |
| [34,35] | ● | CAD | Low | Raytracing | PVWatts | Module level | Module level | From geometry | ○ | ○ |
| [36] | ● | CAD | Low | Transposition | PVWatts | Module level | Module level | From geometry | ○ | ○ |
| [37] | ● | Point cloud | High | Transposition | PVWatts | Module level | Module level | From geometry | ○ | ○ |
| [38] | ● | Point cloud | High | Regression | Regression | Module level | Module level | From geometry | ○ | ○ |

*Module-level Resolution Modeling.* Module-level resolution modeling treats the entire PV module as the basic unit of calculation, standardizing operating conditions such as irradiance and temperature across all cells within a module [28,29]. Unlike cell-level resolution, which models each cell as an individual circuit to capture performance variations of different irradiated solar cells, module-level methods generally employ regression or point-value power models (e.g., SAPM [39] and PVWatts [40]) to estimate the overall module power output. We reviewed the typical module-level modeling frameworks in Table 2, which reveals that some of the frameworks are also incomplete. Similar to the findings for cell-level modeling frameworks, these incomplete modeling frameworks are mainly characterized by a lack of proper methods to incorporate the environmental geometric factors into the computation [30–32], which also prevents shading impacts from consideration.

The complete module-level modeling frameworks can be further classified by the two different sources of geometric input: frameworks that employ CAD models [33–36], and frameworks that employ point clouds [37,38]. Urban point cloud data is primarily collected by airborne scanning with Light Detection and Ranging (LiDAR) devices. Compared to CAD models, point clouds provide richer urban geometric information, and even the geometric features of irregular organisms like vegetations can be well preserved [41], which cannot be accurately modeled in CAD tools. Therefore, with the increase in available urban point cloud data globally [42], modeling frameworks leveraging point clouds would be more advantageous in practical applications, not only because of the ease of access without intensive modeling efforts, but also because the richness of the geometric data in the point clouds helps models to more comprehensively account for the partial shading impacts.

However, by assuming simplified uniform irradiance/shading across the module, module-level modeling frameworks generally provide limited analytical depth in PV performance assessment. Under irregular and dynamic shading situations, module-level frameworks can consistently overestimate power output regardless system inverter configurations compared to high-resolution, cell-by-cell analysis [8].

Table 3. Review of string-level PV modeling frameworks.

| Refs. | Complete framework | Input geometry | Input availability | Irradiance estimation method | Power calculation model | Irradiance resolution | Power resolution | Partial shading definition | Hierarchical System modeling | Support MLPEs |
|---|---|---|---|---|---|---|---|---|---|---|
| [43] | ○ | - | - | Measurement | Efficiency | String level | String level | - | ○ | ○ |
| [44–46] | ○ | - | - | Transposition | Efficiency | String level | String level | - | ○ | ○ |
| [47,48] | ● | CAD | Low | Raytracing | Efficiency | Cell level | String level | From geometry | ○ | ○ |
| [49] | ● | CAD | Low | Transposition | Efficiency | String level | String level | - | ○ | ○ |
| [50] | ● | Point cloud | High | Raytracing | Efficiency | Cell level | String level | From geometry | ○ | ○ |
| [51,52] | ● | Point cloud | High | Transposition | Efficiency | String level | String level | From geometry | ○ | ○ |
| [53,54] | ● | GIS | High | Transposition | Efficiency | String level | String level | - | ○ | ○ |
| [55,56] | ● | Satellite imagery | High | Transposition | Efficiency | String level | String level | - | ○ | ○ |
| [57] | ● | Satellite imagery | High | Transposition | Efficiency | String level | String level | From geometry | ○ | ○ |
| [58] | ● | Satellite imagery & Point cloud | High | Transposition | Efficiency | String level | String level | From geometry | ○ | ○ |

*String-level Resolution Modeling.* String-level resolution modeling approaches take the entire strings or array within a PV system as the basic calculation unit, estimating overall output by assuming identical irradiance and temperature conditions across the system. This approach is generally employed in simplified efficiency-based models (e.g., PVGIS [59] and EnergyPlus [60]) for system output power estimation, differing significantly from the finer analysis scope possible at cell- or module-levels. Such models are particularly useful for large-scale simulations focused on average performance across broad geographic areas, rather than detailed component-specific behaviors [43–46].

String-level modeling frameworks have been widely used to assess the economic and energy potential of widespread PV sectors. As shown in Table 3, complete modeling frameworks at this level often use urban-scale CAD models, point cloud data [47–52], GIS databases that record building characteristics [53,54], or satellite imagery from remote sensing as inputs [55–57]. A state-of-the-art string-level framework that effectively integrates several of these data sources is the one developed by Molin et al. [58]. By automating the identifications of both PV system locations and their individual tilt and orientations, the framework achieved high accuracy, with $R^2$ up to 0.83 when comparing simulated and measured hourly generation for individual systems. Such validated performance makes the string-level approach a reliable choice for scenarios with more uniform conditions and less intricate shading, such as in the simulation of solar parks [61] or in wide-area grid-balancing studies [62].

However, despite the success of well-validated models, the coarse analytical scope of the string-level approach as a category can oversimplify the influence of dynamic environmental variations within urban fibric, inadequately addressing the performance of systems under significant partial shading. Furthermore, the resolution mismatch between detailed irradiance models and simplified power models can also be identified in string-level studies [47,50], which can significantly increases computational effort without guaranteed any improvements in module generation simulation accuracy [8].

### 1.1.2 Effectiveness of MLPEs implementation

The rapid evolution of PV-related technology, particularly Module Level Power Electronics (MLPEs), has introduced new hardware-based dimensions to PV system optimizations [63]. MLPEs are electronic devices configured at the individual module level, and the two representative MLPE technologies - micro inverters and module power optimizers - have gained significant traction in industry. Micro inverters effectively segment the PV system by enabling independent DC-to-AC conversion for each module, ensuring that shading or performance issues affecting one module do not compromise the entire string's output [64]. Module power optimizers, while maintaining the system's central inverter architecture, implement distributed maximum power point tracking (MPPT) at the module level, dynamically optimizing each module's operating point while considering the overall system performance [65]. These advancements have significantly enhanced the adaptability of PV systems to irregular shading patterns, realizing better performance in constrained urban settings [66,67].

The evaluation of MLPEs' effectiveness in mitigating partial shading impacts is the foundation for deciding their application in system design stage. Lee et al. investigates the mismatch loss mitigation effects of MLPEs on a typical rooftop PV array experiencing partial shading from adjacent buildings. The examined PV system consists of two different module orientations, and the results illustrate that the MLPEs prevent annual energy loss by up to 3.23% [68]. Allenspach et al. adopts an analytical method to evaluate the MLPE effects on rooftop PV annual energy production under partial shading from a nearby chimney [69]. The study investigates cases based on different chimney positions and suggests that the electricity gain of MLPE depends on various shading situations. Similarly, Sinapis et al. evaluates the dependence of module power optimizer on various partial shading situations, and highlights the requirements of comprehensive performance modeling approaches that accurately reflect the subtle electrical behaviors and thus assess the feasibility [70].

To effectively model the performance of MLPE-equipped PV systems, a hierarchical modeling approach that captures the electrical behavior throughout the system circuit (i.e., modeling from individual PV components to the entire PV array) is essential. This hierarchical framework enables precise integration of MLPE models at appropriate positions within the PV system architecture, facilitating accurate co-simulation of the power electronics with PV components. Besides, the hierarchical model provides crucial insights into how MLPEs modify the IV characteristics at their implementing level, thus allows engineers to trace performance improvements from the component level to system-wide outputs, providing a comprehensive understanding of how MLPEs optimize power production under various operating conditions. However, from Table 1 to Table 3, it can be found that in the state-of-the-art, only cell-level modeling frameworks can meet the requirement for hierarchical modeling of PV system structures. At the same time, due to the limitation of scalability, none of the existing cell-level modeling frameworks has yet to support the performance evaluation of MLPEs-equipped systems, which highlights an apparent knowledge gap.

### 1.1.3 Summary of findings

Referring to the review on the existing PV modeling frameworks, we found the ones with coarser resolution (module-level or string-level) tend to oversimplify the effects of dynamic partial shading, leading to inaccuracies in predicting system performance. Conversely, modeling frameworks with cell-level resolution will offer in-depth insights into individual cell performance, proving invaluable in urban settings with complex shading patterns. The cell-level resolution approach not only facilitates a hierarchical analysis of micro-level interactions among different system components, but also provides a robust foundation for optimizing system architectures.

However, the cell-level resolution modeling still exhibits limitations in their applicability and scalability. Practical application of existing cell-level approaches is constrained by their dependence on the detailed CAD urban geometry inputs, which are often hard to access and demand large human effort to generate.

Secondly, it is also essential for cell-level modeling frameworks to maintain consistency between the resolutions of irradiance and electrical power models to ensure coherence and accuracy in simulations. Thirdly, a comprehensive demonstration study of the benefits offered by cell-level resolution frameworks with the realistic, urban-specific shading dynamic is lacking, which hinders their attraction to practitioners in real-world scenarios.

Furthermore, existing modeling frameworks designed for conventional PV systems with fixed configurations have difficulties in accommodating and co-simulating modern power electronics like MLPEs. However, insufficient attention and discussion has been given to this scalability issue, highlighting a significant research gap.

In response to these challenges, there is a crucial demand for an improved hierarchical modeling framework for urban distributed PV systems. The improvements should be made to account for (i) nuanced partial shading impacts caused by urban-specific obstructions, (ii) accurate electric behaviors and interactions of different system components within the circuit, (iii) high scalability to various MLPE-equipped configurations, and (iv) high applicability to different modeling scales without intensive human efforts and interventions. These improvements will align with environmental sustainability goals and contribute to addressing the growing societal demand for renewable energy sources.

### *1.2 Research objectives and contributions*

This study addresses the identified knowledge gaps for PV systems performance modeling within urban environments. The research is driven by the following key objectives, which constitute its primary scientific contributions:

- *To develop a novel high-resolution, hierarchical modeling framework for PV systems.* This framework is designed to overcome the limitations of conventional models by accurately simulating performance from the individual solar cell level up to the entire system, with inherent scalability to incorporate various MLPEs.

- *To rigorously validate the proposed framework's predictive accuracy.* This is achieved by systematically comparing the simulated electrical characteristics (I-V-P) and operational states against high-resolution field test data from both conventional PV strings and those equipped with power optimizers under real-world dynamic partial shading.

- *To quantitatively demonstrate the necessity of the high-resolution approach.* The study contrasts the framework's predictions with those from coarser-resolution models to highlight the significant errors and fundamentally incorrect operational states predicted by simpler methods, thereby establishing the importance of cell-level granularity for reliable system analysis in complex urban environments.

- *To contribute a transparent and accessible tool to the research community.* The entire modeling framework has been open-sourced as part of the PYWER Python library[†] (Version 1.0) to promote reproducibility, facilitate future research, and encourage collaborative development in PV systems modeling.

### *1.3 Paper outline*

The paper is structured as follows. Section 2 details the methodology of the proposed modeling framework. Section 3 describes the validation study. In Section 4, we demonstrate the effectiveness of our method by applying the proposed modeling framework to real urban partial shading scenarios. Section 5 discusses the significance of findings, presents its limitations and outlook. Section 6 summarizes the findings and concludes the work.

## 2. High-resolution PV modeling framework

The present study aims at developing an improved PV modeling framework that is able to evaluate PV performance at cell-level resolution, thus satisfying the decision making with different partial shading impacts and MLPE-equipped system configurations. To achieve this, multiple modeling techniques are adapted and expanded in this work, eventually constructing a comprehensive modeling toolchain. Figure 1 describes the PV system performance modeling workflow that is proposed in the present study.

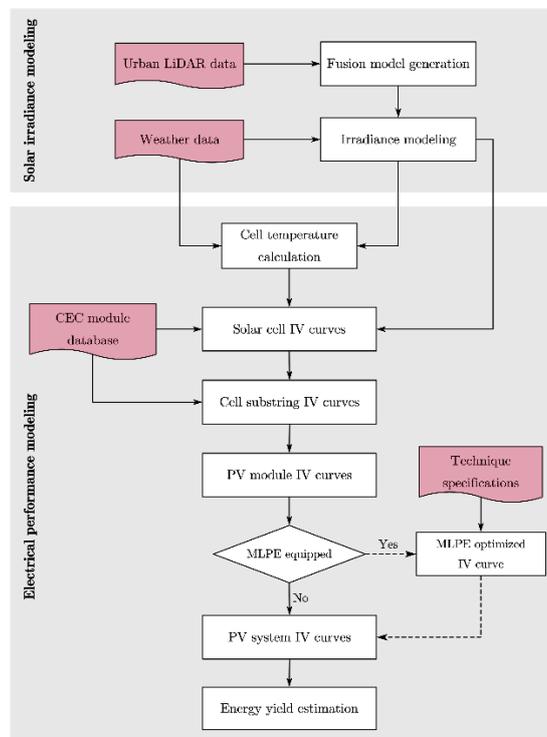

Figure 1. Diagram of PV performance modeling workflow.

---

[†] The developed modeling framework is open sourced at: https://gitlab.tue.nl/bp-tue/pywer

## 2.1 Effective irradiance modeling

The primary step of high-resolution PV performance analysis is the estimation of effective irradiance incident on the solar cells. In urban context, a precise geometric definition of the simulation scene is necessary to enable the irradiance model to account for the potential partial shading impacts. However, as we discussed in Section 1.1.3, the applicability of existing cell-level PV modeling frameworks is limited by the scarcity of detailed urban CAD geometric models. To address this challenge, we employ the innovative "fusion model" developed and validated by Tian et al. [71], which directly takes urban point cloud as input for irradiance calculation. Since high-precision urban point clouds are becoming increasingly available globally, the adoption of this irradiance model can effectively improve the applicability of the PV modeling framework for various urban scenes.

The fusion model employs semantic segmentation methods and surface reconstruction techniques for processing of input point cloud. Specifically, the individual points within an urban point cloud will first be classified into three different categories: vegetation, ground, and buildings. This classification is performed by a well-trained Dynamic Graph Convolutional Neural Network (DGCNN). Subsequently, the classified ground and building points will be reconstructed into precise surface models through Delaunay triangulation techniques, which is reported as the most effective building surface reconstruction algorithm for urban solar potential modeling [26]. This step is crucial for accurately modeling the geometrical characteristics of urban landscapes, which are integral to simulating light reflections and interactions. Vegetation points, however, are preserved in their original form within the model. This decision reflects an understanding of the unique role of vegetation in causing partial shading on PV module surfaces, which cannot be adequately represented by surface models alone. Figure 2 illustrates the generated fusion model for a residential area in Eindhoven, the Netherlands.

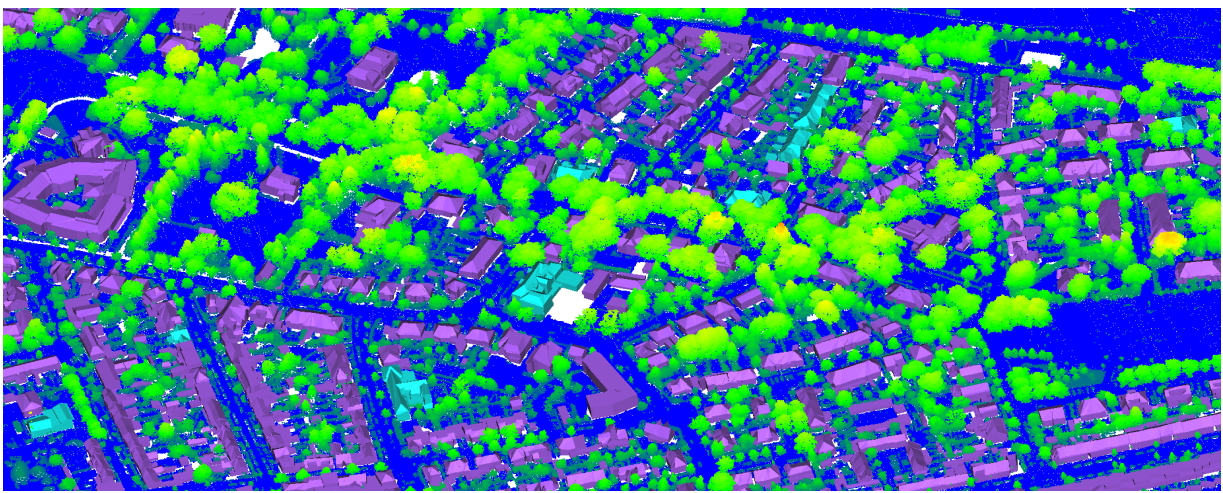

Figure 2. Fusion model of a residential area, with reconstructed buildings in purple, ground in blue, and vegetation point cloud in green.

Within the fusion model, sensor point grids (i.e., one sensor point per solar cell) will be generated over the user-defined PV module locations. To initialize the simulation, individual sensor points will be first situated beneath a virtual discretized sky hemisphere. This conceptual division of the sky into multiple segments, each with an equal solid angle to the centroid, is widely used during the modeling of light dynamics across the urban canopy. Subsequently, at each simulation timestep, the incident irradiance in the area of interest is determined by the following steps:

- Daylight coefficient matrix calculation: Through ray-tracing operations, the model quantifies reflections off regular-shaped urban elements like buildings and the ground. The resulting daylight coefficient matrix $M_{DC}$ [-] embodies the proportion of radiance from each sky segment that reaches predefined sensor points, encapsulating the intricate patterns of irradiance distribution.

- Tree shading ratio matrix calculation: The partial shading impacts of urban vegetation are captured by projecting tree points onto the sky hemisphere. This innovative approach calculates the fraction of area that is covered by the projected points within each sky segment, yielding a tree shading ratio matrix $M_{TSR}$ [-]. This matrix is instrumental in assessing the influence from vegetations on sunlight availability.

- Direct and diffuse irradiance calculation: Subsequently, the solar direct irradiance $E_{direct}$ [W/m$^2$] and sky diffuse irradiance $E_{diffuse}$ [W/m$^2$] is obtained based on the sky radiance matrix $L$ [W/m$^2$] that is generated from weather files (Equation 1 and 2).

- Effective irradiance determination: Effective irradiance $E_{eff}$ [W/m$^2$] describes the amount of irradiance transmitted through the front cover of PV surface under angle-of-incidence effect. We adopted the model by Martin and Ruiz [72] at this final calculation stage, as expressed in Equation 3, correcting the $E_{direct}$ and $E_{diffuse}$ with the angular response correction factor $K(\theta)$, the ultimate summation is $E_{eff}$. The $\theta$ stands for the incident angle of solar beam.

$$E_{direct} = M_{DC} \cdot (1 - M_{TSR}) \cdot L_{direct} \tag{1}$$

$$E_{diffuse} = M_{DC} \cdot (1 - M_{TSR}) \cdot L_{diffuse} \tag{2}$$

$$E_{eff} = E_{direct} \cdot K(\theta) + E_{diffuse} \cdot K(60°) \tag{3}$$

As discussed in Section 1.1, for a PV performance modeling framework, it is important to align the irradiance model's resolution with the subsequent electrical power model. The cell-level irradiance data from the fusion model approach ensures a seamless integration with the cell-level PV electrical performance model that will be described in follow-up sub-sections, particularly ideal for urban settings where shading and reflections significantly impact irradiance levels.

## 2.2 PV cell temperature modeling

PV operation temperature also plays a role in the photovoltaic conversion process, as an operating temperature increase above 25 °C negatively affects the electrical efficiency of PV modules [73]. In the design stage or when there is a lack of direct measurement data, PV cell temperature can be modeled based on the weather data and the thermal properties of the PV module itself. Readers can refer to [74] for a comprehensive review of the PV temperature modeling strategies in the state-of-the-art.

In the present study, we adopted the transient cell temperature model by Fuentes [75]. The outstanding performance of this model has been validated in multiple prior experiments, with a prediction accuracy achieved 95% [75,76]. In comparison with other temperature models, Fuentes model also presents consistent lowest prediction RMSE under both clear sky and cloudy sky conditions [77].

The Fuentes model is predicated on the principle of energy balance, positing that the solar energy incident on a PV module is primarily allocated between conversion to thermal energy and electrical energy. The thermal portion contributes to the module's temperature and is modulated by environmental interactions, such as convective and radiative heat losses. Concurrently, the electrical portion is harnessed and directed away through the external circuit. The mathematical explanation of Fuentes model is provided in Appendix A.1.

## 2.3 Electrical performance modeling

Referring to the modeling workflow in Figure 1, the PV electrical performance will be modeled based on the obtained effective irradiance and temperature. Figure 3 illustrates the structure of a PV system with crystalline-silicon modules, the major components involved are highlighted and denoted separately. Strictly following the hierarchical structure of the PV system, the present study also models the PV system energy performance hierarchically. The proposed method starts from modeling the electrical behaviors of the smallest component (the individual solar cell), then building up through cell substrings and modules (Figure 3a), and finally, module strings and the entire PV system (Figure 3b). This layered methodology allows for detailed analysis and understanding of system behavior under a wide range of operational conditions.

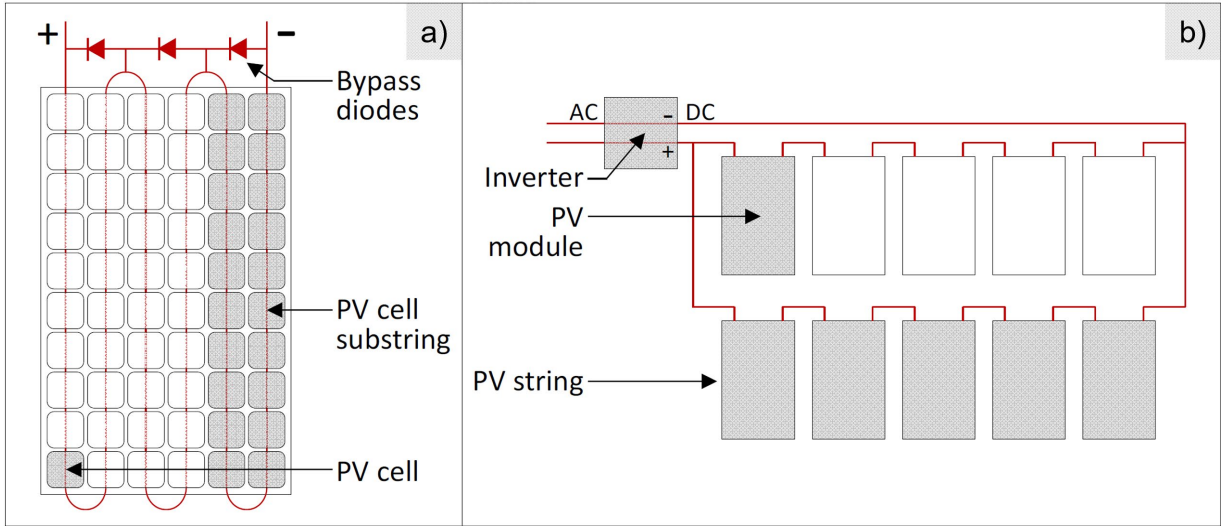

Figure 3. Schematic diagram of a) PV system consisting of two 5 module long strings connected in parallel to a string inverter, and b) a 60-cell crystalline-silicon PV module consisting of three cell substrings. The named components are highlighted in gray. Adapted from [78] and modified by the authors.

### 2.3.1 PV cell modeling

As illustrated in Section 1.1, by modeling the PV cell as equivalent electric circuit, the n-diode models provide estimation of electrical behavior and IV curves. Among the models, the one-diode model provides a simplified representation of PV cell behavior, while the two-diode model introduces an additional diode to account for recombination losses both within the depletion region and in the quasi-neutral regions [79], thus enhance the prediction accuracy across a broader spectrum of operating conditions. Built upon this, the present methodology adapts the two-diode model for cell IV curve calculations (Figure 4) [80].

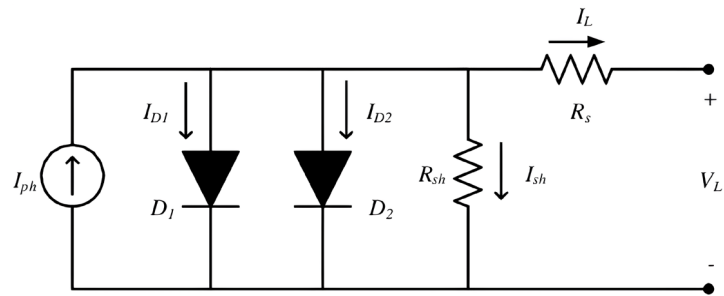

Figure 4. Two-diode model of solar cell.

The circuit consists of a photovoltaic source with photogenerated current $I_{ph}$ [A], two parallel diodes, a series resistance ($R_s$) and a shunt resistance $R_{sh}$ [Ω]. Mathematically, the terminal voltage and current of the PV cell can be related as:

$$I = I_{ph} - I_{D1} - I_{D2} - \frac{V + I \cdot R_s}{R_{sh}} \tag{4}$$

where $I_{D1}$ and $I_{D2}$ are the currents [A] through the first and second diodes, calculated as:

$$I_{D1} = I_{sat1}\left(e^{\frac{V + I \cdot R_s}{n_1 \cdot V_t}} - 1\right) \tag{5}$$

$$I_{D2} = I_{sat2}\left(e^{\frac{V + I \cdot R_s}{n_2 \cdot V_t}} - 1\right) \tag{6}$$

here, $I_{sat1}$ and $I_{sat2}$ are the saturation currents [A] of the diodes, $n_1$ and $n_2$ are the ideality factors [-] that reflect the recombination mechanisms within the cell, while $V_t$ is the thermal voltage [V].

To solve the two-diode model equation, instead of using the constant parameter values, we followed the approach outlined by Ishaque et al. [81] with an automatic acquisition strategy of model parameters to guarantee the modeling accuracy. We present detailed mathematical derivation and implementation procedures for the solution in Appendix A.2.

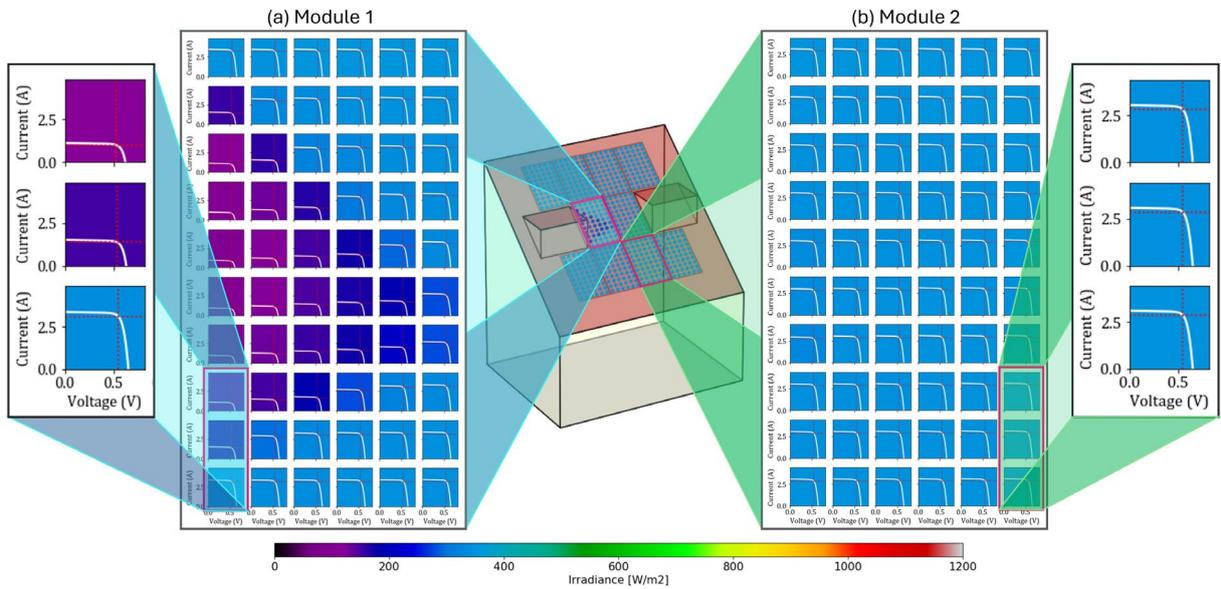

Figure 5. Visualization of modeled solar cell IV curve within (a) partially shaded module and (b) unshaded module.

Figure 5 visually illustrates the model's capability in capturing partial shading effects on solar cell IV curves. The example focuses on a south-facing, roof-mounted PV system with a 45-degree tilt, which is subject to partial shading from nearby dormer windows. The specific conditions shown are for 16:00 on April 24, 2022, where the shading impact is obvious. Two adjacent modules are selected for zoomed-in inspection.

Figure 5a illustrates Module 1, which is partially covered by a "triangular" shadow cast by a dormer. This creates a highly non-uniform irradiance profile across the module surface, with values ranging from approximately 50 W/m² in the deepest shade to 370 W/m² in the fully sunlit portions. The resulting solar cell IV curves reflect this heterogeneity; compared to the unshaded cells, the shaded cell curves indicate significantly lower current levels across the entire voltage spectrum, implying a pronounced reduction in electrical output. In contrast, Figure 5b depicts Module 2, which is completely unshaded at this time. It receives a uniform irradiance of 370 W/m² across its surface. Consequently, the IV curves for the individual cells within this module are virtually identical, showcasing the ideal, homogeneous operating condition.

### 2.3.2 From cell to system

The transition from single-cell modeling to broader system-level analysis is essential for accurately predicting overall PV performance under various working conditions. By starting with the fundamental IV characteristics of an individual cell, we can sequentially build up substrings, modules, and eventually entire PV strings or systems [82]. This hierarchical approach - as highlighted by Karatepe et al. [83] - allows us to systematically capture mismatch effects, incorporate bypass diode behavior, and account for other practical considerations that become significant as multiple cells and modules are interconnected. A detailed description of how substring-, module-, and string-level IV curves are derived from individual cell characteristics can be found in Appendix A.3.

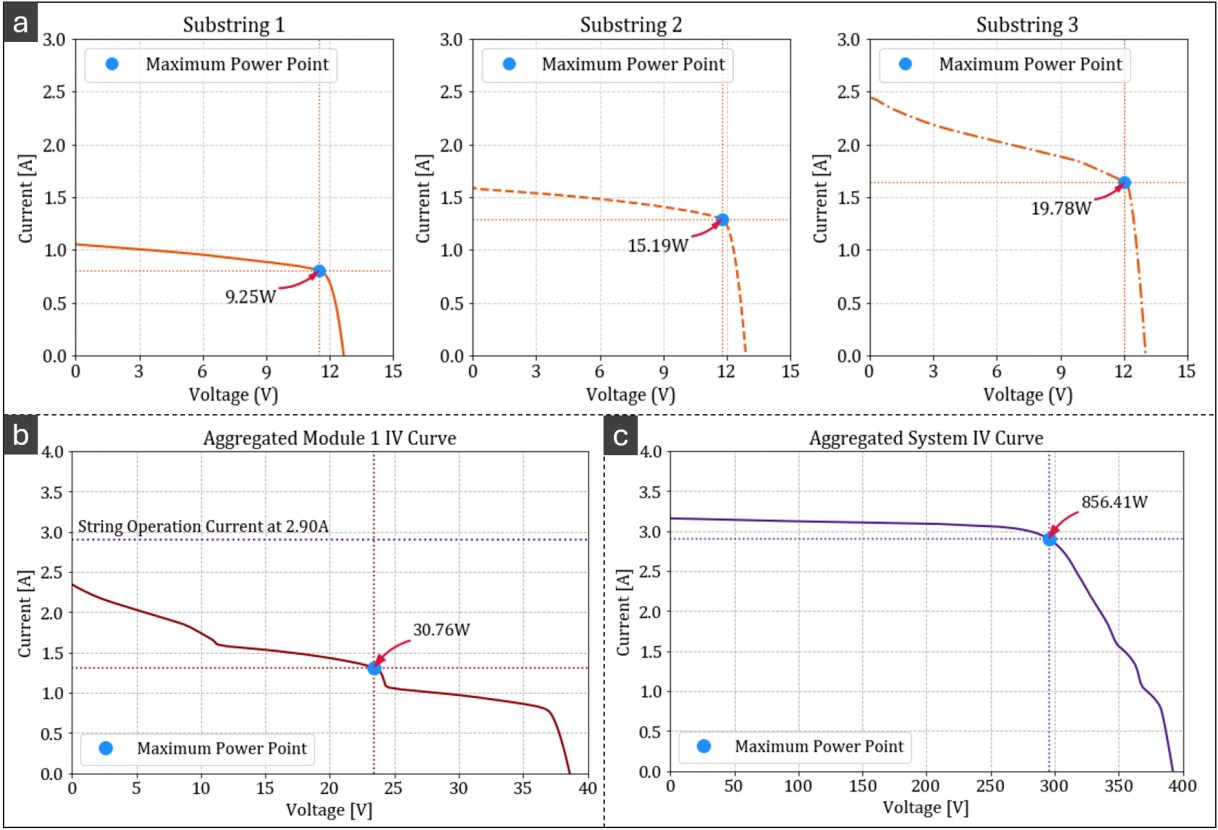

Figure 6. Visualization of hierarchically aggregated IV curves of different PV system layers, with (a) cell substrings, (b) individual module, and (c) entire system.

Figure 6 illustrates this hierarchical IV curve aggregation approach. To highlight the propagation of mismatch effects from individual solar cells to the system performance, we take the heavily shaded Module 1 in Figure 5 as an example. Figure 6a displays the calculated IV curves for the three series-connected cell substrings within Module 1. Due to varying exposure to the dormer's shadow, their electrical characteristics diverge significantly. Substring 1, being the most heavily shaded, exhibits a severely limited short-circuit current of approximately 1.1A and a Maximum Power Point (MPP) of only 9.25W. In contrast, the least-shaded Substring 3 achieves a short-circuit current of 2.5A and an MPP of 19.78W. The performance of Substring 2 falls between these two extremes, clearly demonstrating how shading differences across a single module create distinct electrical behaviors at the substring level.

Subsequently, these substring characteristics are integrated to derive the module-level IV curve, as shown in Figure 6b. The resulting curve is less smooth and features obvious knees, which are characteristic of aggregating mismatched series-connected substrings. The module's own MPP is calculated to be 30.76 W (at approximately 23.7 V and 1.3 A). However, at this specific time, the overall PV string is operating at a current of 2.90A. Since this operating current is significantly higher than Module 1's own short-circuit current (approximately 2.40A), all three of its internal bypass diodes are activated. Consequently, the entire module is bypassed, contributing no power to the system and highlighting how severe mismatch can completely neutralize a module's energy production.

Finally, Figure 6c presents the IV curve for the entire 10-module PV system. The distortions caused by the partial shading on Module 1 are still observable as minor knees on the system-level curve. However, because the partial shading is localized to a single module in this scenario, the system's overall MPP, as regulated by the central inverter, occurs at a point on the curve before these shading-induced impacts become dominant.

### *2.4 Module level power electronics modeling*

Recognizing the gap in existing PV system modeling tools in Section 1.1.2, which cannot effectively evaluate the operational characteristics and benefits conferred by MLPEs, the present study introduces extended capabilities for MLPEs modeling. Specifically, two additional models have been developed - one for micro inverters and another for power optimizers - allowing the users to seamlessly incorporate them into the purposed high-resolution PV system modeling backbone for co-simulations.

The developed microinverter model encapsulates the core operational principle of DC-AC converters with MPPT. By performing the conversion at the module level, the system can mitigate power losses from partial shading and module mismatch, leading to improved energy performance under non-uniform conditions [64]. For the simulation, the model determines the microinverter's harvested power by identifying the MPP directly from the detailed module IV curve calculated for each timestep.

Similar to a microinverter, the power optimizer model begins by tracking the MPP of the individual PV module. However, as a DC-DC converter, it then focuses on managing power flow within the DC part of the system. The developed model is able to simulate the two distinct operating modes of buck power optimizers:

- *Conductive Mode*: When the current demanded by the central inverter for the entire string is lower than or equal to the module's available current at its MPP ($I_{string} \leq I_{mpp}$), the optimizer enters a pass-through state. In this mode, it acts similarly to a series-connected component with minimal intervention, allowing the module's operating point to be dictated by the central inverter's system-level MPPT.

- *Buck Mode*: When the string current demanded by the inverter is higher than what the module can supply at its MPP ($I_{string} > I_{mpp}$), which is common for a shaded module in a string of unshaded ones, the optimizer actively engages. In this mode, the optimizer's MPPT ensures the module operates at its own MPP to harvest the maximum available power.

By actively enabling the buck mode and decoupling the module's operation from the central inverter's regulation, the optimizer prevents the shaded module from being forced to a sub-optimal, low-power operating point. Readers can refer to the work by Erickson and Maksimovic [84] for detailed explanation of DC-DC converters, we also provided the theoretical details of the developed power optimizer model in Appendix A.4.

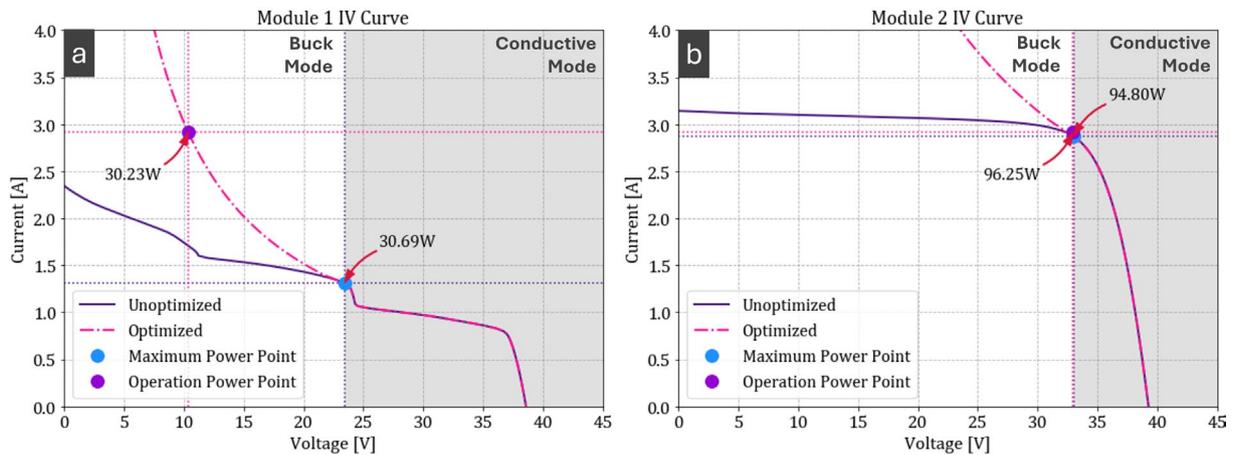

Figure 7. Visualization of module-level IV curve with and without power optimizer application, wherein (a) partially shaded Module 1 and (b) unshaded Module 2.

To visualize the MLPE modeling results, Figure 7 continues the analysis of the two representative modules from the demonstration system previously discussed in Figure 5. Each subplot displays two different IV curves. The first is the module's original unoptimized IV curve, on which the tracked MPP is highlighted to represent the modeled operating point that would be selected by a microinverter. The second is the optimized IV curve, which represents the effective characteristic at the power optimizer's output, with its specific operating power point (OPP) annotated.

Comparing these curves reveals the optimizer's distinct modes of operation. For operating points to the right of the module's MPP (where voltage is high and current is low, i.e., $I_{string} \leq I_{mpp}$), the optimized curve aligns closely with the module's original curve. This region represents the optimizer's passive "conductive mode," where it has minimal intervention. Conversely, to the left of the MPP ($I_{string} > I_{mpp}$), the optimized IV curve significantly deviates from the module's curve. Here, the current level is substantially elevated across the voltage spectrum, representing the active "buck mode," where the optimizer lets the module work at its MPP and converts the harvested power to meet string requirements.

Figure 7a details the performance of Module 1, which is subjected to significant partial shading. In this scenario, the string current is considerably higher than the module's own MPP current, forcing the optimizer to operate in "buck mode." The optimizer decouples the module from the string, allowing it to operate at its true MPP of 30.69W. However, due to the optimizer's internal conversion losses, the final power delivered at the optimizer's output OPP is slightly lower, at 30.23W.

In contrast, Figure 7b shows the results for the unshaded Module 2. Because it is not suffering from shading, its MPP current is slightly higher than the string current, causing the optimizer to operate in "conductive mode." In this case, the module's OPP is dictated directly by the string conditions and is located on the near right side of its MPP. This results in a final output power of 94.80W, which is very close to its available MPP of 96.25W, indicating minimal losses while in this pass-through mode.

## *2.5 Energy production calculations*

Having clear understandings of the IV relations within the PV system, we are able to determine the energy output $P(t)$ [W] at a specific timestep $t$ by performing MPPT. Subsequently, the instantaneous energy yield $Y_{inst}(t)$ [Wh] can be calculated by Equation 7, wherein the $\Delta t$ stands for the duration of the timestep [h].

$$Y_{inst}(t) = P(t) \cdot \Delta t \tag{7}$$

To obtain an integrated view of the energy production capacity of the PV system over a given timeframe, we can further calculate the cumulative energy yield $Y_{cum}$ [Wh] by Equation 8, wherein $N$ is the total number of timesteps within the period, and $Y_{inst}(i)$ is the instantaneous yield of the PV system at timestep $i$.

$$Y_{cum} = \sum_{i}^{N} Y_{inst}(i) \tag{8}$$

# 3. Validation

To evaluate the accuracy and robustness of the proposed high-resolution PV performance modeling methodology, this section presents a detailed validation study comparing the minute-wised simulation results against data from a comprehensive field test work [70]. The validation is structured to first assess the model's accuracy for a conventional PV system and then to evaluate its capability in modeling the performance of modules equipped with power optimizers under both unshaded and partially shaded conditions.

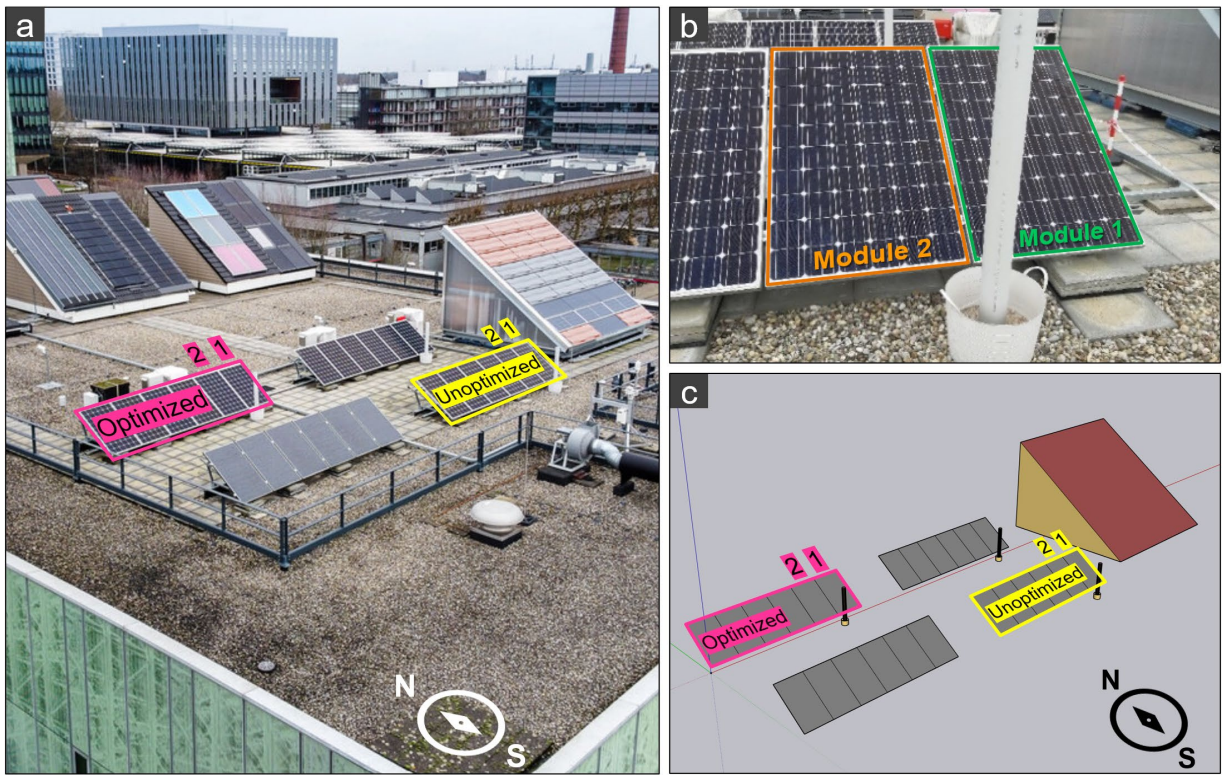

Figure 8. Field test setup of validation study, wherein (a) overview of the validation scene; (b) pole obstruction for creating partial shading on module surface; and (c) geometry definitions for validation simulations.

The field measurement campaign was conducted on the campus of Eindhoven University of Technology (TU/e) on the roof of the "Vertigo" building, as depicted in Figure 8. The measurement system consists of two individual PV strings with 6 Yingli Panda 265Wp modules, each string connected to a separate 1.5 kW inverter.

As highlighted in Figure 8a, the string on the front (south) side serves as a reference "Unoptimized" system, without any module-level power electronics. Conversely, the string on the back (north) side is the "Optimized" system, where module-level power optimizers are selectively applied to the first two modules in the string (Module 1 and Module 2, with numbering from right to left). The one-minute resolution DC electrical parameters of each PV module in both systems were monitored by a calibrated

Yokogawa WT1800 high-performance power analyzer, which provides a high power accuracy of ±0.1%.

To validate the PV modeling accuracy under the influence of partial shading, two identical pole shading elements, each with a height of 146 cm and a diameter of 12.3 cm, were utilized during the field test. As shown in Figure 8b, the poles were placed at the exact same position relative to Module 1 and Module 2 in both the unoptimized and optimized strings to ensure that both systems operated under identical, reproducible shading conditions. The corresponding geometry model of this validation scene, which serves as the direct input for the modeling framework, is shown in Figure 8c.

### 3.1 Reference module performance

The modeling framework's accuracy for a conventional PV system is first validated against field test data from the "Unoptimized" reference string. Figure 9 compares the simulated and measured operating voltage for Module 1 and Module 2 on April 20th, 2019. This clear-sky day was selected specifically for its pronounced partial shading conditions, allowing for a rigorous assessment of the model's dynamic performance.

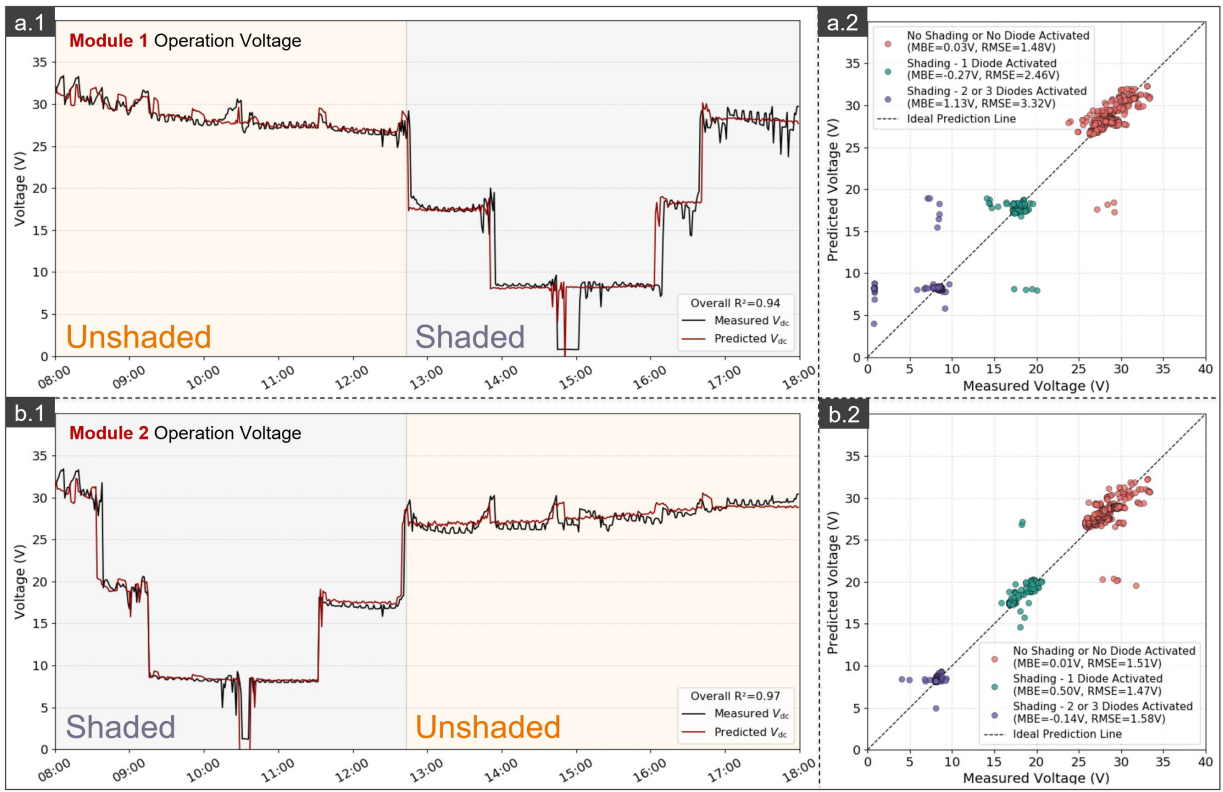

Figure 9. Comparison of predicted reference modules' operation voltage against field measurements on April 20th, 2019.

The validation results for the reference Module 1 are shown in Figure 9a. The time-series plot (Figure 9a.1) illustrates a strong agreement between the predicted (red) and measured (black) operating voltage dynamics throughout the day. The model successfully captures both the relatively stable voltage during unshaded periods and the frequent, significant voltage drops caused by bypass diode activations during partial shading. This excellent fit is quantitatively reflected by an overall $R^2$ score of 0.94. Some slight temporal mismatches can be observed, where sudden voltage changes are predicted approximately five minutes earlier (e.g., 14:55-15:00, 16:00-16:05) or later (e.g., 14:45-14:50) than measured. These minor discrepancies are attributed to the model determining cell shading based on the solar position relative to each cell's centroid; in reality, a shadow edge may trigger a bypass diode before the cell's center is covered. Nonetheless, these edge cases do not significantly affect the overall model performance.

The scatter plot in Figure 9a.2 further highlights the model's accuracy by categorizing the data points based on the activation status of the module's three bypass diodes. The majority of points are concentrated around the ideal prediction line. For conditions with zero or one bypass diode activated, the prediction achieved remarkable small Mean Bias Errors (MBE) (at 0.03V and -0.27V), with Root Mean Squared Error (RMSE) at 1.48V and 2.46V, respectively. When two or three diodes are active, the prediction error increases slightly to an MBE of 1.13V and an RMSE of 3.32V, which is attributable to the temporal mismatches during these rapid transitions. However, this slightly increased error value remains sufficiently low to be acceptable for high-resolution PV performance modeling tasks.

Similarly, Figure 9b presents the validation results for Module 2. The time-series data in Figure 9b.1 shows a near-perfect alignment between the predicted and measured voltage variations, resulting in a robust overall $R^2$ score of 0.97. This promising performance is confirmed in the scatter plot (Figure 9 b.2), where data points from all categories closely follow the ideal prediction line. For Module 2, the absolute MBE is consistently at or below 0.50V, and the RMSE remains stable around 1.5V across all bypass diode activation states. This stability further implies the model's robust accuracy.

Overall, the results presented in Figure 9 confirm the high accuracy and reliability of the modeling framework in predicting the operational voltage of standard PV modules, correctly capturing the complex dynamics introduced by partial shading and bypass diode behavior.

### *3.2 Operation points for MLPE applications*

After evaluating the proposed modeling framework's performance for the conventional PV system, the second stage of validation assesses its ability to predict the input operating point for MLPE applications. This involves tracking and predicting the module's MPP characteristics - power, voltage, and current - and comparing them against the measured data. As described in Section 2.4, this MPP represents the direct power conversion point for a microinverter and the target input for a module power optimizer.

The validation results for Module 2 within the "Optimized" string are presented in Figure 10. Specifically, Figure 10a to 10c present the set of results for MPP power, voltage, and current, respectively. Each set includes a time-series line chart comparing the predicted and measured data throughout the day (subplots a.1, b.1, c.1), alongside a scatter plot quantifying the model's accuracy (subplots a.2, b.2, c.2). For the scatter plots, data points are categorized by whether they occurred during the partially shaded period or the unshaded period.

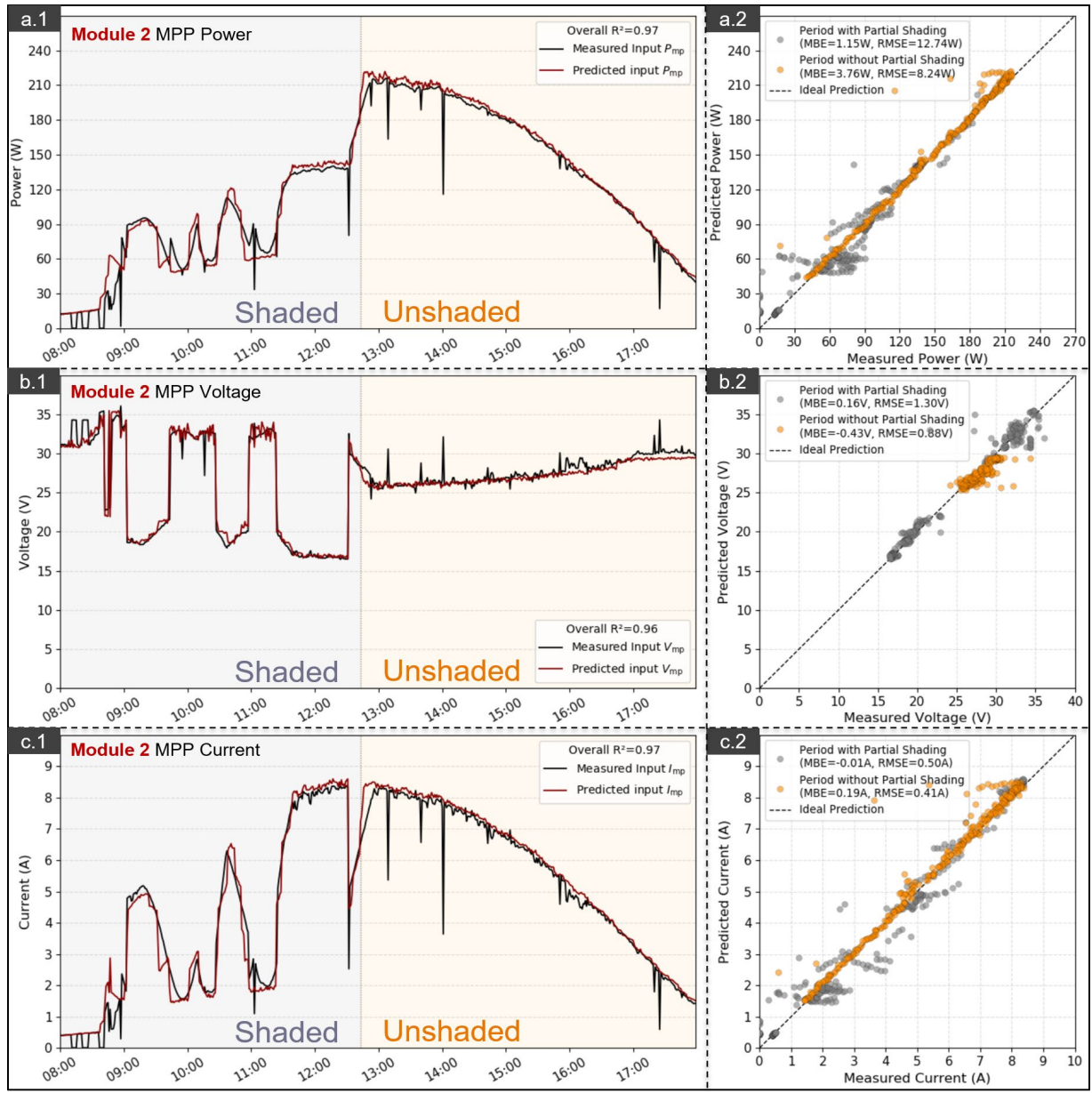

Figure 10. Comparison of predicted module-level MPP characteristics against field measurements for MLPE operations on April 20th, 2019.

From the time-series charts, the model's predictions show very high agreement with the field measurements for most of the day. With overall R² scores for power, voltage, and current predictions at 0.97, 0.96, and 0.97, respectively, these results confirm a high degree of accuracy in predicting the module's MPP characteristics under dynamic solar irradiation and temperature conditions. It is worth noting that some sudden, sharp drops in measured power and current (with corresponding voltage spikes) during the unshaded period (after 12:45) are not reflected in the model's predictions. These instances are attributed to transient measurement events where the power analyzer's MPPT algorithm may have briefly tracked a local, rather than global, MPP, and are not indicative of a model deficiency.

A detailed analysis of the scatter plots reveals the model's performance under different conditions. During the partial shading period (before 12:45), the majority of data points in all three plots are concentrated around the ideal prediction line. The MBE for MPP power, voltage, and current is exceptionally low, at 1.15W, 0.16V, and -0.01A, respectively, implying minimal systematic error under complex shading influences. However, some visible deviations can be found, which increase the RMSE during this period to 12.74W, 1.30V, and 0.50A, respectively. As seen in the time-series plots, these deviations arise because the measured MPP characteristics form a relatively smooth curve even during shading transitions, while the model's predictions, being calculated independently for each one-minute time step, exhibit more high-frequency fluctuations. This difference in "smoothness" can be attributed to the inherent operational characteristics of real-world MPPT systems compared to the instantaneous nature of the simulation. A physical MPPT algorithm in a power analyzer or inverter has a non-zero settling time [85,86]; it continuously perturbs the operating point and requires a brief period to hunt for and stabilize at a new MPP after a change in conditions [87]. This, combined with the electrical capacitance of the physical system [88], and potential data averaging within the logger [89], results in a smoothed or dampened response to rapid fluctuations in the module's true MPP. In contrast, the simulation framework calculates a static MPP for each discrete one-minute interval, leading to sharper transitions between operating points.

Finally, for the unshaded period, the data points converge even more tightly to the ideal prediction line, with the RMSE values for power, voltage, and current decreasing to 8.24W, 0.88V, and 0.41A, respectively. The visible outliers in this category are again attributed to the previously mentioned transient measurement events.

### 3.3 Output performance of module power optimizer

As the final stage of the validation study, the output of the module power optimizer model is compared against field test data to confirm the framework's accuracy for MLPE-equipped systems. This stage evaluates the combined performance of the PV module model and the subsequent power optimizer model. The predicted output results for Module 2 in the "Optimized" string are presented against measurements in Figure 11. The results from Figure 11a to 11c are detailed for power, voltage, and current, respectively. Each scatter plot (subplots .2) now visualizes the power optimizer's output points (in red) alongside the input MPP points from the previous validation stage (in green) for direct comparison.

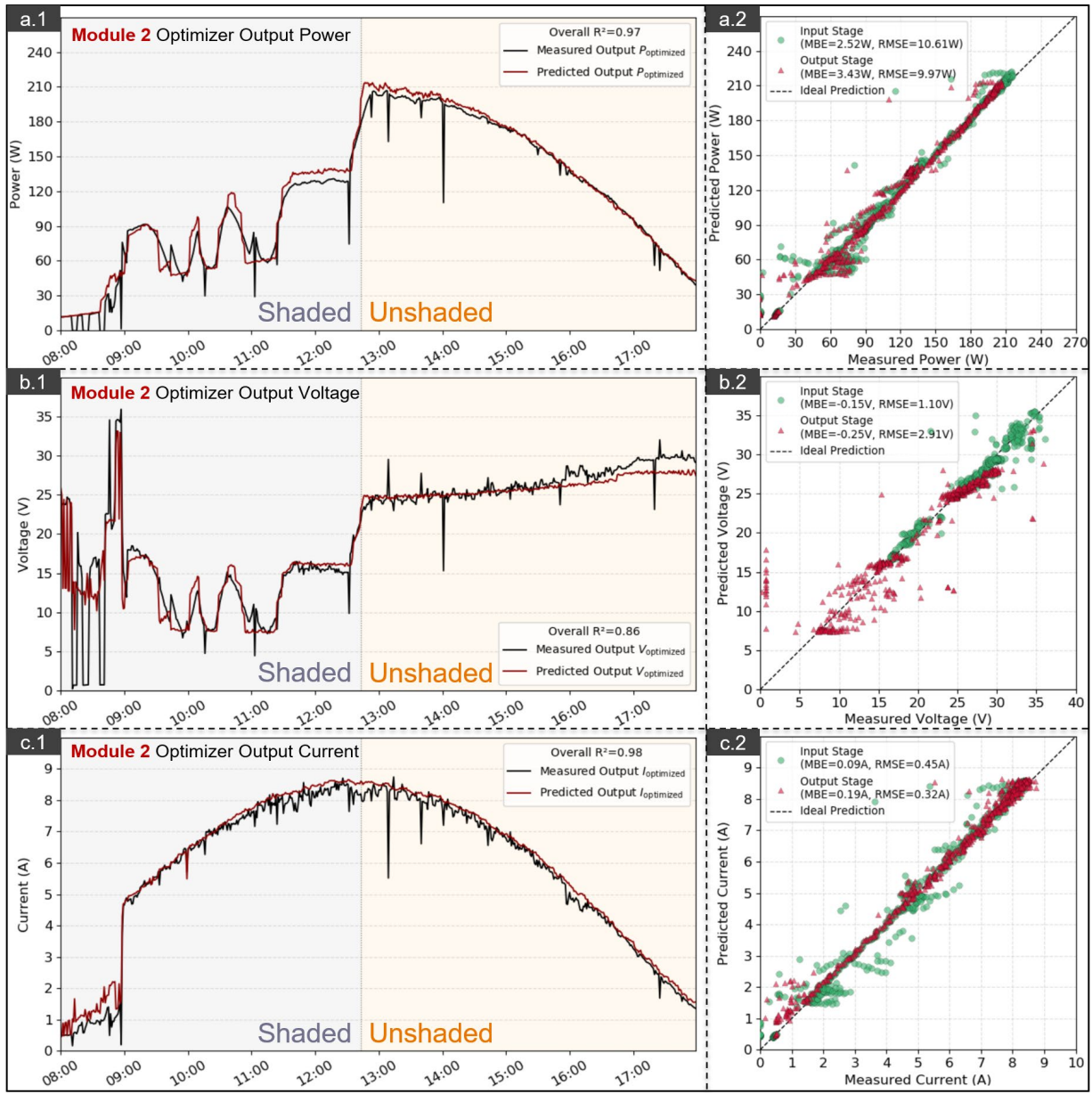

Figure 11. Comparison of predicted module power optimizer output characteristics against field measurements on April 20th, 2019.

From the time-series line charts, the model's prediction of the optimizer's output aligns well with the field test results in both trend and value for most of the day. For the investigated output power (Figure 11a.1) and current (Figure 11c.1) predictions, the model achieved consistently high $R^2$ scores of 0.97 and 0.98, respectively. The scatter plots for power (Figure 11a.2) and current (Figure 11c.2) further confirm this, showing that the output predictions closely follow the ideal prediction line, with MBE and RMSE values that are very similar to the input stage results.

For the output voltage predictions (Figure 11b.1), the overall $R^2$ score experiences a decrease to 0.86 compared to the input stage. This is caused by the extreme voltage fluctuations measured during the morning low-irradiance period (before 08:55), which are not fully captured by the model. These mismatches subsequently cause some output data points in the lower-left corner of the scatter plot (Figure 11b.2) to be distributed with more variance, resulting in an amplified MBE (from -0.15V to 0.25V) and RMSE (from 1.10V to 2.91V) compared to the input stage. The reason for these measured voltage bounces relates to the inherent behavior of the real-world system at its operational limits. In the early morning, the available power from the entire PV string is extremely low, often near or below the minimum operational threshold of the central inverter. In this state, the inverter's MPPT algorithm struggles to find a stable lock and may perform wide voltage sweeps or repeatedly attempt to start up, resulting in the observed extreme fluctuations. The simulation, however, predicts a stable operating point based on the complete IV curve for each instant, rather than reproducing these rapid hardware-level control dynamics during low-irradiance conditions. Nevertheless, given that the photovoltaic current is substantially low during this period (Figure 11c.1), these voltage prediction mismatches result in only marginal power prediction errors (≤10W, Figure 11a.1) and do not significantly affect the overall prediction accuracy during primary production hours.

Additionally, the characteristics presented in Figure 11 highlight the power optimizer's effectiveness. During the partial shading period (before 12:45), the output current level (Figure 11c.1) is significantly increased compared to the input current level shown in Figure 10, while the voltage level (Figure 11b.1) is correspondingly suppressed. This clearly indicates the optimizer was activated and operating in "buck mode," elevating the output current to match the string's requirements and thus mitigating mismatch loss. Conversely, during the unshaded hours, the output current and voltage remain at similar levels to the input results, indicating the optimizer adaptively switched to its pass-through "conductive mode." These behaviors are further confirmed in the scatter plots, where the red output voltage points (Figure 11b.2) show a clear downward shift relative to the green input points, while the output current points (Figure 11c.2) are concentrated in a higher region. The inevitable power conversion loss is also visible in Figure 11a.2, where the red output power points are distributed slightly below the green input points.

## 4. Demonstration study

After validating the proposed high-resolution modeling framework, a demonstration study was developed to further demonstrate its effectiveness and practical application, focusing on a real-life residential dwelling in Eindhoven, the Netherlands. The objective is to explore the model's performance in predicting the PV system's electrical behavior under more complex, realistic shading scenarios. As shown in the overview in Figure 12a, the demonstration is carried out for a large two-story dwelling where a PV system containing 24 modules, with a total installed capacity of approximately 6.36kWp, is to be deployed on the south-facing roof area. Notably, this setup faces complex partial shading from architectural elements like dormer windows, as well as dynamic shading from nearby roadside trees.

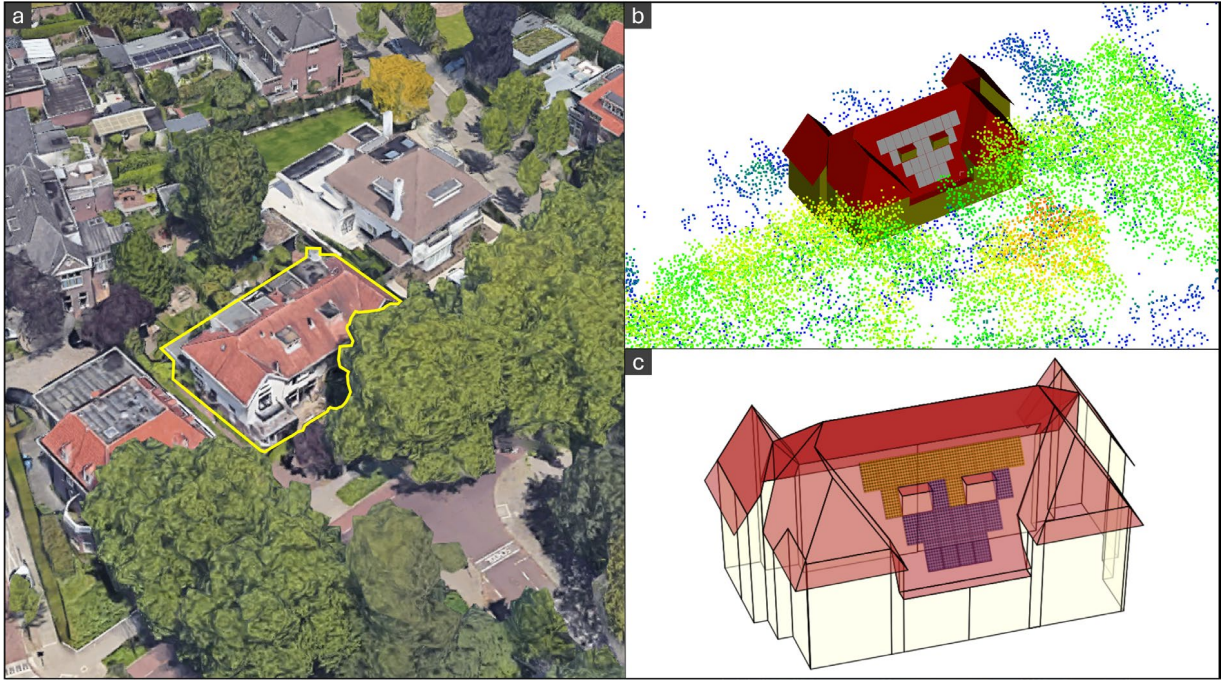

Figure 12. A dwelling with a rooftop PV system under design that is subject to complex partial shading impacts, with (a) aerial imagery of the scene, (b) generated fusion model of the scene for irradiance simulation, and (c) string configuration of the system, with the heavier shaded String 1 is in purple, and less shaded String 2 in orange.

The proposed approach is used to predict the system's performance at different irradiation moments, with a focus on evaluating the uncertainty introduced by different modeling resolutions. The simulation inputs are derived from urban point cloud data sourced from the Netherlands AHN3 terrestrial information database [90] and historical weather data for Eindhoven in 2022 obtained from KNMI weather stations [91]. Figure 12b shows the fusion model used for the irradiance simulation, which accurately reproduces the spatial location of the studied dwelling, the PV system, and the surrounding tree geometries. To minimize losses from energy mismatches among modules in the design phase, the methodology proposed by McNeil [35] was adopted to configure the PV modules into two strings based on their annual cumulative irradiation exposure (Figure 12c). String 1 (purple) includes the modules most severely affected by shading, whereas String 2 (orange) consists of modules with less shading impact, each string connects to a separate 3kW inverter.

A key aspect of this demonstration is the examination of variances in IV curve results between the proposed high-resolution method and other coarser-resolution approaches commonly used in practice. The difference between these approaches can be specifically understood as follows: In the high-resolution simulation, individual solar cells within a module are each equipped with a unique sensor point, allowing for the direct utilization of point-specific irradiance and temperature data in the IV curve calculations. Conversely, coarser resolutions aggregate the irradiation and temperature conditions over larger areas. This study investigates three such resolutions: substring-level, which assigns the average irradiance of sensor points within a substring to all *n* cells in that substring (Equation 9); and similarly for the module and string levels (Equations 10 and 11), where *m* and *l* represent the number of substrings per module and modules per string, respectively.

$$E_{eff\_substring} = \frac{1}{n} \sum_{i=1}^{n} E_{eff\_cell,i} \qquad (9)$$

$$E_{eff\_module} = \frac{1}{m} \sum_{j=1}^{m} E_{eff\_substring,j} \qquad (10)$$

$$E_{eff\_string} = \frac{1}{l} \sum_{k}^{l} E_{eff\_module,k} \qquad (11)$$

### *4.1 PV module IV curve estimation*

To comprehensively understand the dynamic shading influence within the demonstration scene, we start by analyzing the individual PV module IV curve modeling results at different spatial resolutions. Two partially shaded modules, designated Module A and Module B (positioned in different strings) were selected for detailed analysis at four individual moments (09:00, 10:00, 11:00, and 12:00) on April 24[th]. The simulated irradiance distribution across the PV array for cell-level resolution modeling are provided in Figure 13, which confirms the performance of Module A is significantly affected by dynamic partial shading from nearby trees throughout this period, while Module B is primarily impacted by shading from a dormer window. To maintain content coherence and manage article length, details on irradiance distributions at coarser resolutions are provided in Appendix B.1.

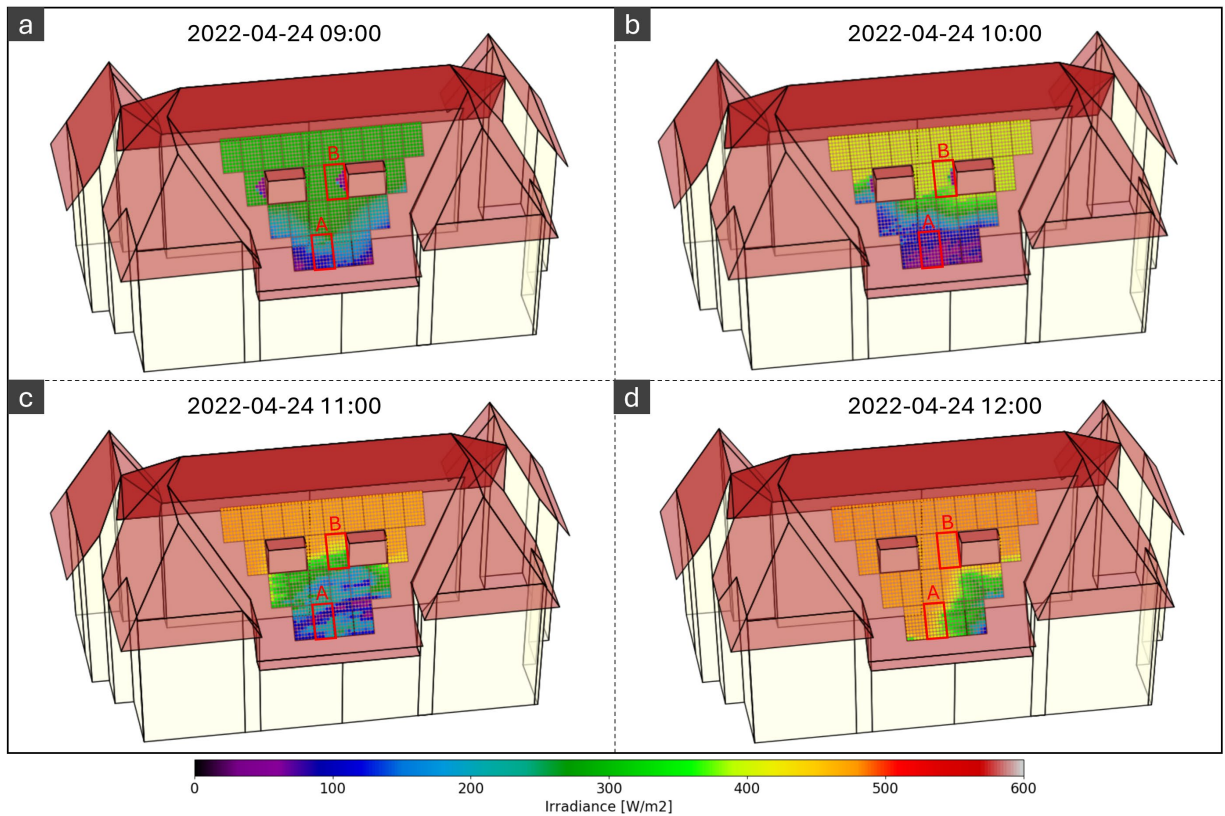

Figure 13. Effective irradiance distribution modeled by cell-level resolution at four different timestamps, where the two modules for data analysis are highlighted by red bounding boxes with annotation A and B.

The modeling results for Module A at the four investigated timestamps are presented in a grid format in Figure 14, where each row (a-d) corresponds to a specific timestamp. Within each row, the columns visualize different aspects of the simulation: the irradiance distribution at cell-level (column .1), substring-level (column .2), and module-level (column .3) resolutions; the resulting module IV curves with annotated Maximum Power Points (MPPs) for each resolution (column .4); and the same IV curves showing the module's actual Operating Power Points (OPPs) within the string (column .5).

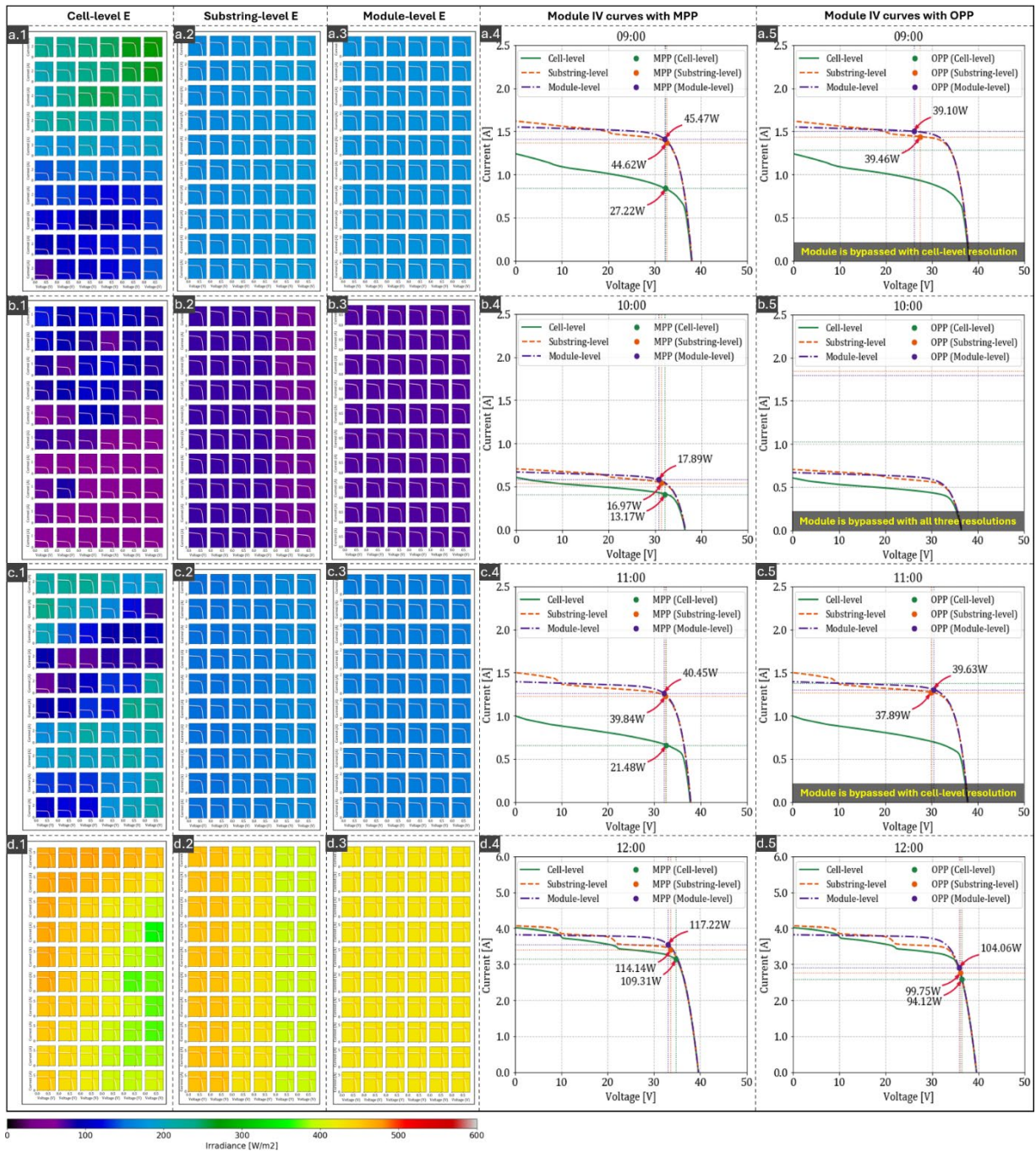

Figure 14. Modeling results of Module A at four individual timestamps on April 24th, 2022.

A detailed analysis of the results reveals the substantial impact of partial shading and the critical importance of modeling resolution. At 09:00, significant shading from the tree canopy is evident. The cell-level model (Figure 14a.1) captures a complex pattern where the most shaded cells receive only around 100 W/m² irradiance, while less shaded cells receive up to 280 W/m². This highly uneven irradiance distribution causes severe power mismatch losses, which is reflected in the cell-level IV curve showing substantial current reduction and a low MPP of 27.22W (Figure 14a.4). Furthermore, the analysis of the module's operation within the string (Figure 14a.5) confirms that under these conditions, Module A is fully bypassed and contributes no power to the string yield. In contrast, the coarser substring-level and module-level resolutions (Figure 14a.2, 14a.3) fail to capture the detailed shadow pattern, instead simplifying the irradiance across the module surface to a uniform average of about 190 W/m². This oversimplification leads to a significant misrepresentation of the shading impact, with the IV curves for these resolutions showing higher currents and MPPs around 45W (Figure 14a.4), an overestimation of approximately 65% compared to the cell-level result. Critically, as shown in Figure 14a.5, these coarser models incorrectly predict that the module is still contributing around 39W to the system, completely failing to identify its bypassed state. Similar conclusions can be drawn from the 11:00 results (Figure 14c), where the two coarser resolutions overestimate the module MPP by 86% and again fail to predict that the module is bypassed.

For the other two investigated timestamps, the shading complexity over Module A is reduced. At 10:00 (Figure 14b.1), the module is almost fully shaded, while at 12:00 (Figure 14d.1), the partial shading influence is mitigated, resulting in less pronounced irradiance differences among cells (≤100 W/m²). Consequently, the IV curves from the coarser resolutions appear relatively close to the cell-level results. However, overestimation persists; the module's MPP is still overestimated by up to 36% at 10:00 and 7% at 12:00 by the coarser models. In terms of operational status, the model correctly identifies that the module is bypassed by all resolutions at 10:00 due to the severe shading (Figure 14b.5). Conversely, at 12:00, all resolutions correctly predict that the module is working normally, but the specific OPP is still overestimated by 6% (substring-level) and 11% (module-level), as shown in Figure 14d.5.

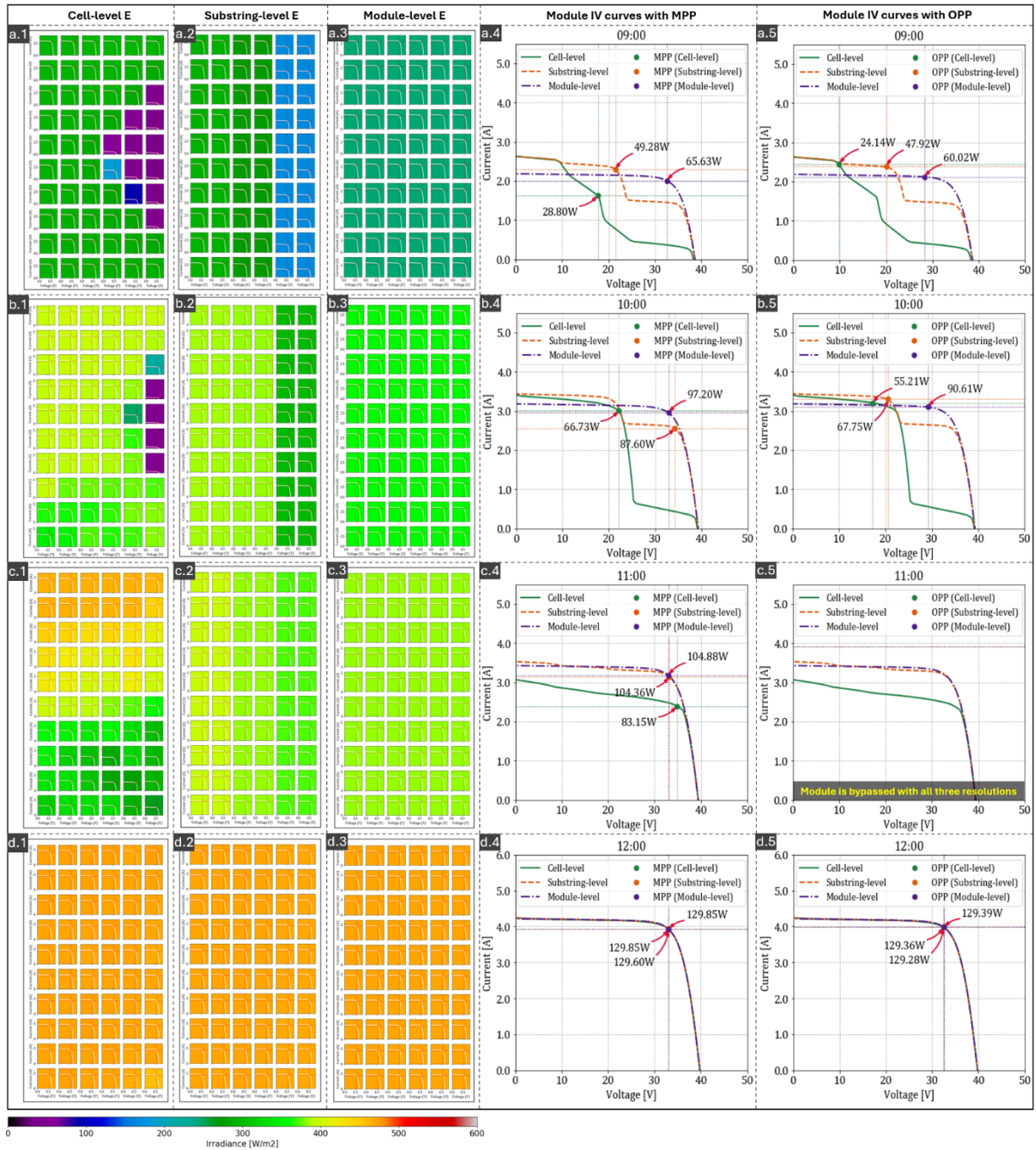

Figure 15. Modeling results of Module B at four individual timestamps on April 24th, 2022.

The analysis continues with Module B, whose performance is detailed in Figure 15. At 09:00, the cell-level irradiance results (Figure 15a.1) confirm that the rightmost solar cell substring is severely shaded by the dormer window, with heavily shaded cells receiving significantly reduced irradiance of around 70 W/m². Consequently, the cell-level resolution IV curve displays noticeable knees caused by current drops, resulting in the lowest calculated MPP at 28.80W (Figure 15a.4) and an OPP of 24.14W (Figure 15a.5). However, the coarser resolution models show significant deviations. The substring-level resolution aggregates this into less detailed, bar-shaped shadows (Figure 15a.2), while the module-level resolution further simplifies the incident irradiance to a uniform average of about 240 W/m² (Figure 15

a.3). These simplifications propagate substantial errors into the IV curve predictions. The substring-level model overestimates the MPP by 71% (at 49.28W) and the OPP by 99% (at 47.92W). The error is further enlarged by the module-level resolution, which overestimates the MPP and OPP by 128% (65.63W) and 149% (60.02W), respectively. At 10:00 (Figure 15b), although the partial shading influence on the rightmost substring is mitigated, the IV curve results show a consistent trend where both the MPP and OPP are still notably overestimated by the two coarser resolution models.

The shading source for Module B switches to the roadside trees at 11:00. Due to the semi-transmissive nature of tree canopies, the irradiance reductions are less severe, but the distribution becomes more chaotic, as captured by the cell-level model in Figure 15c.1. This scenario is analogous to the conditions observed for Module A at the same timestamp (Figure 14c). The two coarser resolutions largely ignore this complexity, simplifying the irradiance to stable values around 380 W/m² (Figure 15c.2, 15c.3). As shown in Figure 15c.4, these simplifications result in nearly identical IV curves with MPPs around 104W, an overestimation of approximately 25% compared to the more accurate cell-level MPP of 83.15W. Furthermore, because the rest of the modules in String 2 are less shaded at this time, the string current regulated by the central inverter is substantially higher than Module B's short-circuit current. As a result, Figure 15c.5 confirms that the module was correctly predicted to be bypassed across all three resolutions.

Finally, at 12:00, Figure 15d shows an absence of noticeable shading, leading to a uniform irradiance distribution across the module. In this unshaded condition, the IV curve estimations and calculated MPPs converge across all resolutions, demonstrating that the models agree when complex partial shading is not a factor.

*4.2 PV string IV curve estimation*

In the demonstration scenario, the 24 modules are configured into two strings based on their annual cumulative irradiation (Figure 12). This allows for an exploration of the aggregated electrical behavior of multiple modules under varied shading conditions. The IV curve modeling results for heavily shaded String 1 and less shaded String 2 at 10:00 and 11:00 are presented in Figure 16. For this string-level analysis, alongside the cell-, substring-, and module-level resolutions, a string-level resolution is included. This common, coarser approach assumes uniform irradiance across all modules in the string, facilitating a broad comparison of how spatial resolution affects system performance assessment.

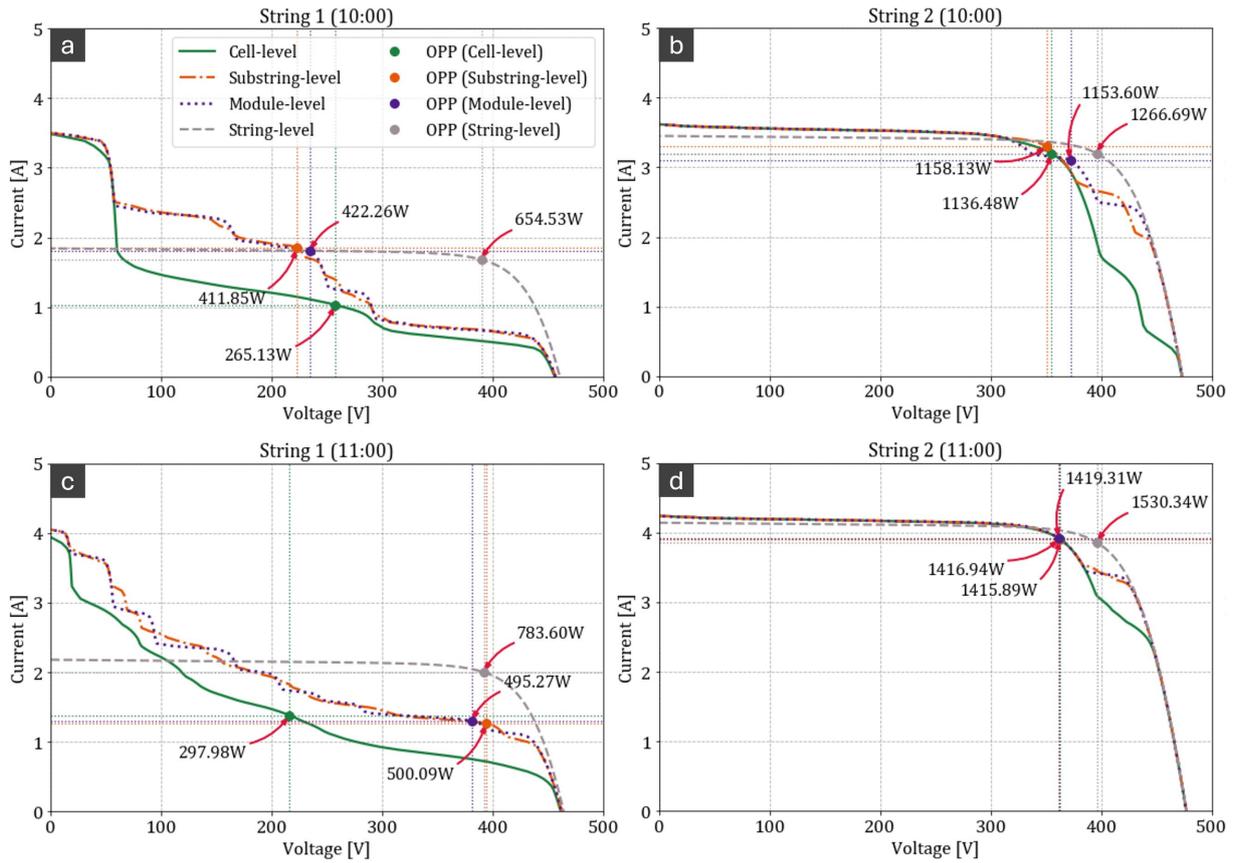

Figure 16. String IV curves that modeled by different resolutions, where (a) string 1 at 10:00, (b) string 2 at 10:00, (c) string 1 at 11:00, and (d) string 2 at 11:00.

For String 1, which suffers substantial partial shading influence, the noticeable current drops (knees) in its aggregated IV curve are the direct manifestation of performance mismatch among series-connected modules. The cell-level resolution model is able to accurately capture these mismatch effects, which are critical for determining the string's real-time operational behavior. As shown in Figure 16a, the cell-level curve showed a significant current drop from ~3.4A to ~1.8A at a string voltage of approximately 50V, this limited string currents ultimately yielding a suppressed OPP of only 265.13W. In contrast, the substring- and module-level models, while identifying the current drop at a similar voltage, fail to capture its full magnitude, underestimating the severity of the mismatch effects and predicting a healthier IV curve. This finding is consistent with the per-module overestimations observed in Figure 14, as these errors propagate and aggregate to the string level. Consequently, these coarser models predict erroneous OPPs at 411.85W and 422.26W, overestimating the string's actual power potential by up to 59%. For the string-level resolution, which assumes uniform irradiance, a smooth IV curve is that completely ignores these mismatch effects is generated, leading to a drastically incorrect OPP prediction of 654.53W - roughly 147% higher than the more realistic cell-level estimate. This trend persists at 11:00 (Figure 16c), where the cell-level model predicts a significant mismatching IV curve with OPP of 297.98W, while the substring- and module-level models overestimate it by about 68%, and the string-level model overestimates it by a remarkable 163%.

In contrast, as String 2 experiences minimal shading, its IV curve has a clearer, well-defined maximum. As a result, the OPPs predicted by the finer resolutions (cell, substring, and module) are all in close agreement. At 10:00 (Figure 16b), the predicted OPPs fall within a tight range of 1136.48W to 1158.13W. By 11:00 (Figure 16d), they converge around 1417W - an output approximately 375% higher than the heavily shaded String 1 at the same moment, highlighting the effectiveness of the irradiance-based stringing strategy. Although the cell-level IV curve exhibits steeper current drops at high voltages (beyond 380V), this detail correctly captures the behavior of the few slightly shaded modules but does not affect the overall power prediction, as the string's OPP occurs well before this voltage drop-off. However, even under these minimally shaded conditions, the string-level resolution remains inadequate, overestimating the string OPP by approximately 10% at 10:00 and 8% at 11:00 compared to the finer resolutions.

Similar trends at 9:00 and 12:00 are provided in Appendix B.2 to avoid repetition. Overall, these results highlight the effectiveness of high-resolution modeling strategy in capturing accurate system operation power dynamics under shading influences.

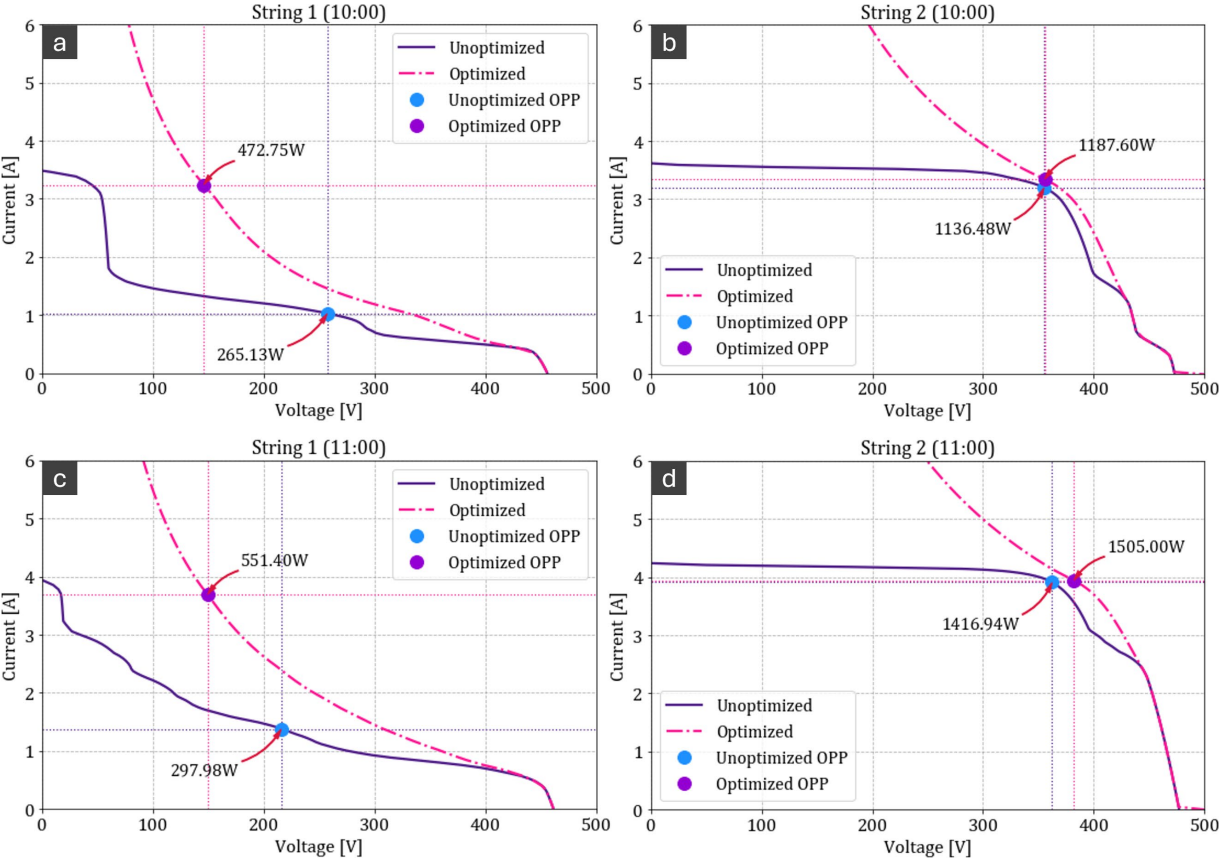

Figure 17. String IV curves with and without power optimizer equipped that modeled by different resolutions, where (a) string 1 at 10:00, (b) string 2 at 10:00, (c) string 1 at 11:00, and (d) string 2 at 11:00.

Building on the high resolution (cell-level) performance analysis of the demonstration strings, we then evaluate the effects of adopting module power optimizer. Figure 17 contrasts the modeled IV curves with OPPs of the original unoptimized strings (in purple) with the optimizer-equipped cases (in pink). As shown, the application of power optimizers results in significantly steeper aggregated string IV curves. This is because optimizers equipped on partially shaded modules operate in buck mode, dynamically converting each module's MPP to an optimal string feed-in point with elevated current and bucked voltage. The more modules that are partially shaded, the more significant this power optimization effect becomes at the string level.

The performance improvement is most pronounced for the heavily shaded String 1. At 10:00 (Figure 17 a), the optimized OPP is elevated by a remarkable 78%, from 265.13W to 472.75W. An even larger gain of 85% is observed at 11:00 (Figure 17c), with the OPP increasing from 297.98W to 551.40W. Conversely, for the less shaded String 2 (Figure 17b, 17d), the power gains are marginal, with improvements of only 4% and 6% at 10:00 and 11:00, respectively. However, in these less shaded cases, the system will still benefit from the optimizers' ability to mitigate hotspot effects and potential long-term degradation during intermittent shading [92].

The effectiveness of the optimizers stems from the operational flexibility they provide to the central inverter. By decoupling the modules, the inverter is free to regulate the string operating current to a level that maximizes the output of the unshaded or most productive modules, without being limited by the low current of a single shaded module. This is also clearly demonstrated in Figure 17. For the heavily shaded String 1, the optimizers operate predominantly in buck mode, significantly elevating the string's operating current to align with that of the more productive String 2. For instance, at 10:00 (Figure 17a), the OPP current for String 1 is boosted from approximately 1.0A to 3.2A, closely matching the ~3.3A operating current of String 2 (Figure 17b). This alignment persists at 11:00, with String 1's current elevated to ~3.8A, near String 2's ~3.9A (Figure 17c, 17d). In contrast, since the majority of modules in String 2 remain unshaded, its unoptimized operating current is already high. Consequently, the optimizers on String 2 remain mostly in "conductive mode," and the optimized OPP current is nearly identical to the unoptimized one.

Furthermore, we also performed calculations for the optimized string IV curves at 9:00 and 12:00, and the results were supplemented in Appendix B.3, from which we can obtain the same conclusions as in Figure 17.

### *4.3 Monthly string energy production estimation*

To demonstrate the proposed modeling framework's ability to predict long-term cumulative energy yield, the system's operation was simulated over a full year, with the energy production decomposed by month. This analysis aims to reveal the impacts from both modeling resolutions and MLPE application on annual yield assessment. To assist data interpretation, we calculated the direct sky clearness index ($KT_{dir}$ [-]), which denotes the ratio of direct beam radiation on a horizontal plane to the extraterrestrial horizontal radiation [93] and infers the prevailing sky conditions across the months. A higher $KT_{dir}$ indicates clearer skies with more direct solar irradiance, which can exacerbate partial shading effects.

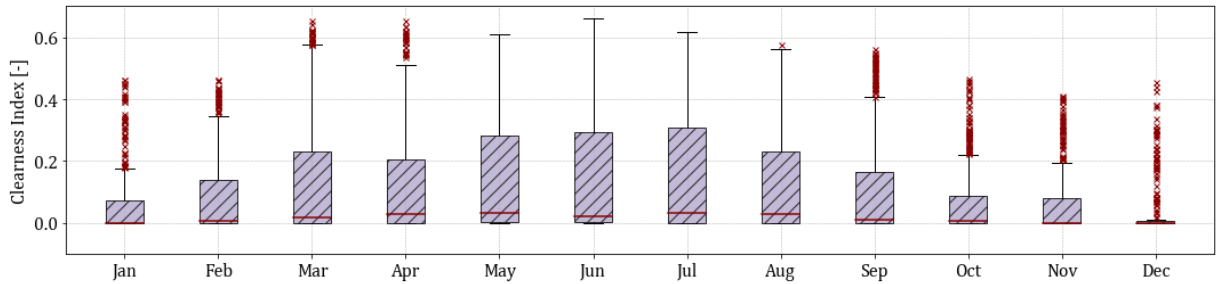

Figure 18. Distribution of hour-wise direct sky clearness index in different months throughout the year 2022.

The boxplots in Figure 18 segregate the hourly $KT_{dir}$ during daylight hours for each month. The lower and upper bounds of the boxes represent the first and third quartiles of the clearness index, respectively, with the median value marked by a red line. The period from March to September shows higher median and third quartile values, indicating frequent clear-sky conditions conducive to pronounced partial shading events. Conversely, the lower values from October to February indicate prevalent overcast conditions, reducing the severity of partial shading with sharp shadows due to diminished solar irradiance. This seasonal pattern aligns with the expected solar path and mild maritime climate of the location, where lower solar elevations and frequent cloud cover during winter reduce direct solar radiation.

### 4.3.1 Modeling resolution impacts

The impact of spatial modeling resolution is first evaluated by comparing the monthly energy yield predictions for the two investigated PV strings, with the results presented in Figure 19. The observations directly reflect the seasonal weather patterns shown in Figure 18. During the period with a high clearness index (March to September), significant discrepancies emerge among the different modeling resolutions. For both strings, the monthly energy yield predicted by the substring-, module-, and string-level resolutions consistently show notable overestimations (⩾10 kWh) compared to the cell-level results. For instance, in March, the substring- and module-level models overestimate energy production by approximately 25% for the heavier shaded String 1 (Figure 19a) and 16% for the less shaded String 2 (Figure 19b). Meanwhile, the string-level resolution resulted in more drastic discrepancies, particularly for String 1, with overestimations up to 54% in March and ranging from 16% to 45% across April through September. These significant prediction errors observed in the coarser models are directly attributable to their inability to accurately model severe mismatch losses under strong, direct sunlight, as detailed in the IV curve analysis (Figure 16).

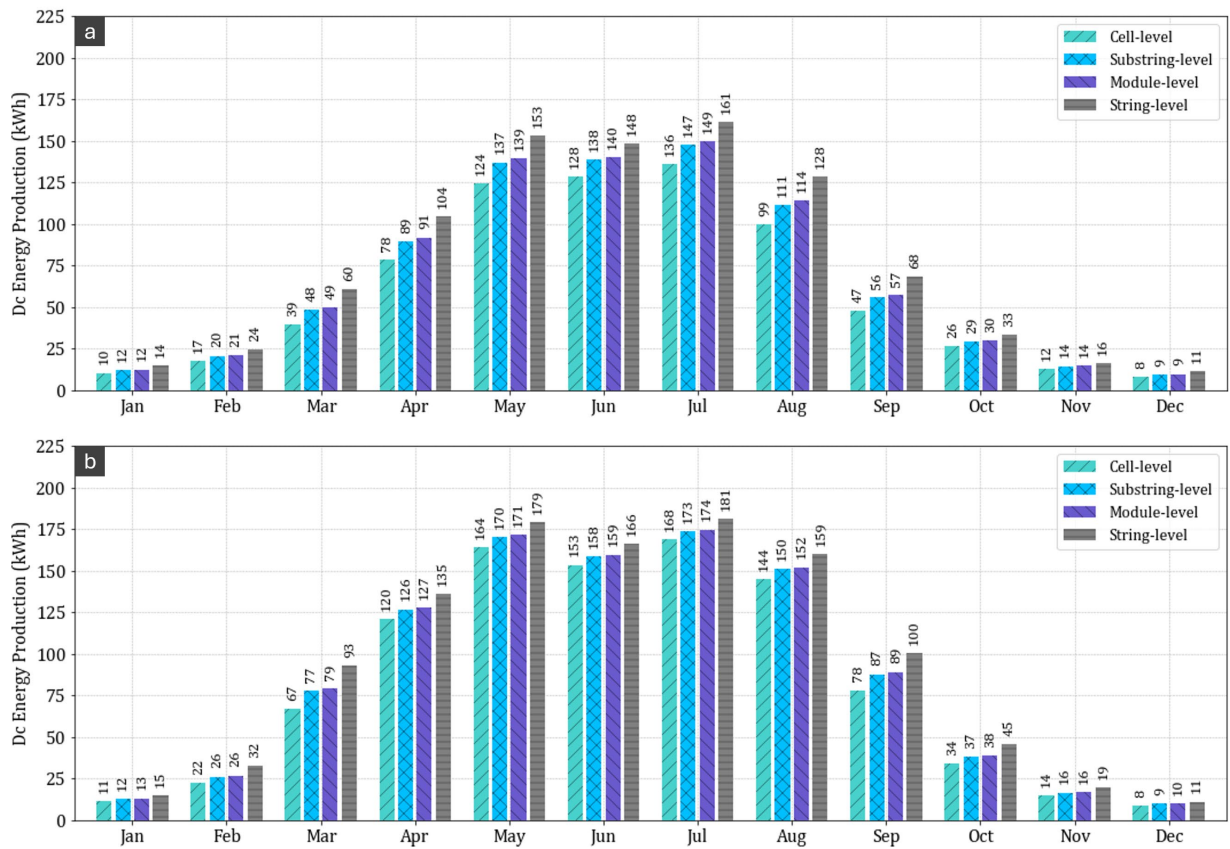

Figure 19. Modeled monthly DC energy production by different resolutions of (a) String 1 and (b) String 2.

In contrast, during the months from October to February, the discrepancies among the resolution results are minimal for both strings. The energy yield overestimation is typically under 5 kWh per month for the substring- and module-level models and under 10 kWh for the string-level model compared to the cell-level prediction. This convergence aligns with the lower sky clearness index observed in Figure 18, which indicates a dominance of diffuse irradiance that softens partial shading effects. Furthermore, as illustrated by the seasonal shading patterns in Figure 20, the lower solar path during winter months causes the nearby trees to cast a more extensive and uniform shadow across the entire PV array. When all modules are more uniformly shaded, the mismatch between modules is reduced, causing the predictions from different resolutions to naturally converge.

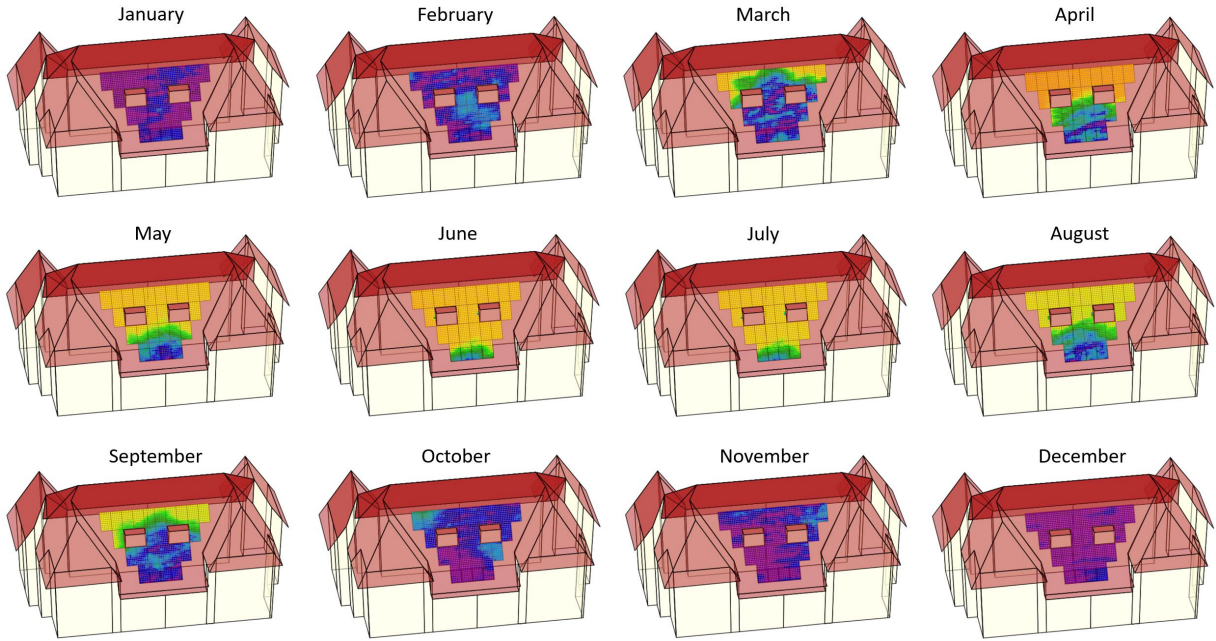

Figure 20. Irradiance distribution over the demonstration PV array at 11:00 on typical clear days of each month.

Overall, this monthly analysis quantitatively demonstrates that the necessity for high-resolution, cell-level modeling is not uniform throughout the year but is most critical during seasons with a high potential for dynamic partial shading. While coarser models may suffice for overcast conditions or periods of uniform shading, they introduce significant errors in energy yield prediction for systems susceptible to complex, direct-beam shading scenarios.

### 4.3.2 MLPE application impacts

Subsequent analysis focuses on the potential effects of MLPEs on improving the energy performance of the demonstration system. Specifically, two reference MLPE-free scenarios (regular system operated by string or central inverter) and two MLPE-equipped scenarios (with power optimizers or microinverters) are evaluated. The effectiveness of MLPEs is quantified by comparing the monthly string energy production of MLPE-equipped scenarios against the reference ones, as presented in Figure 21. It is important to note that these results represent the DC-side energy yield to highlight the performance gains from mismatch mitigation, before accounting for DC-AC inversion losses.

Consistent with the findings from the resolution analysis (Figure 19), from October to February, the monthly energy production differences between the reference scenarios, as well as between the MLPE scenarios are minimal ($\leq 2$ kWh). The energy uplift provided by the MLPEs is also significantly suppressed during the period - compared to the string inverter scenario, the monthly improvements are less than 4 kWh for power optimizers and 7 kWh for microinverters. This aligns with the environmental data (Figure 18, Figure 20), which indicates that the predominantly overcast sky conditions and extensive, uniform tree shading reduce the inter-module irradiance variations, thus diminishing the effectiveness of per-module optimization. This suggests that in climates or seasons where diffuse conditions or uniform full-string shading prevail, the added value of MLPEs may be limited and not justify the investment.

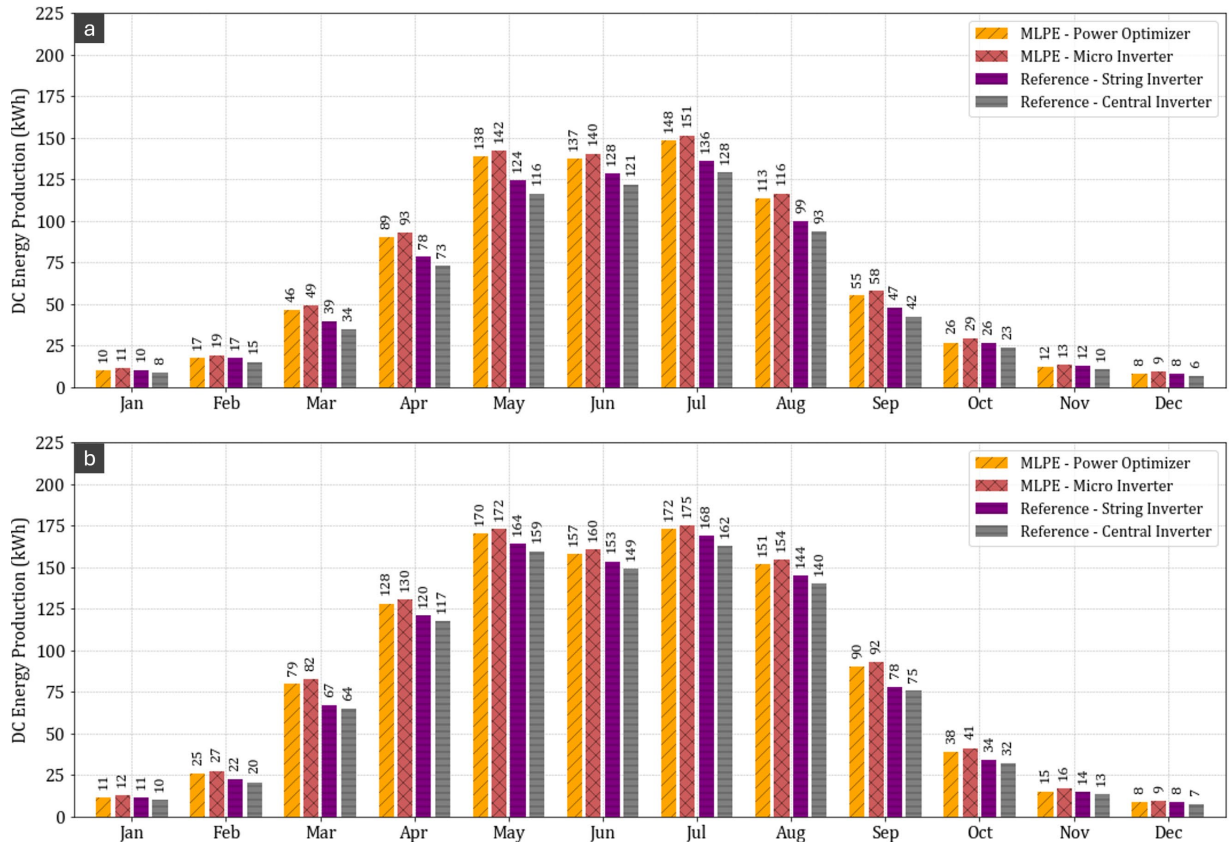

Figure 21. Modeled monthly DC energy production of the reference MLPE-free scenarios and MLPE-equipped scenarios for (a) String 1 and (b) String 2.

Conversely, from March to September, the effectiveness of MLPEs becomes significant. For the heavily shaded String 1 (Figure 21a), which consistently experiences noticeable partial shading, the peak energy gain from MLPEs occurs in May. The power optimizer system yields 138 kWh, around 14 kWh (11%) improvement over the string inverter scenario and around 22 kWh (19%) improvement over the central inverter scenario; while the microinverter scenario demonstrates more apparent gains, with an energy yield of 142 kWh, which is 18 kWh (15%) and 26 kWh (22%) higher than the string and central inverter scenarios, respectively. Additionally, although less affected by shading, String 2 (Figure 21b) still benefits notably from MLPE applications, with energy enhancements in March reaching up to 15 kWh compared to the string inverter scenario and 18 kWh compared to the central inverter scenario. Moreover, the magnitude of these energy enhancements varies seasonally. The gains from MLPEs are more pronounced in the "shoulder" months (e.g., March-May, August-September) than in the peak summer months (June, July). This is because the higher solar positions in mid-summer led to shorter shadow paths and less time under partial shading influence, as visualized by Figure 20.

Throughout the March-September period, the microinverters consistently outperform the power optimizers by a margin of approximately 3 kWh per month for both strings. This slight performance advantage can be attributed to their fundamental architecture; microinverters maximize the power of each module independently by immediately converting its DC output to AC, which completely isolates it from the performance of other modules. In contrast, while power optimizers also perform per-module MPPT, their DC outputs are pooled into a series string that must operate at a common current. This additional DC-DC conversion step and the constraint of conforming to a string-level operating point can result in slightly higher curtailment and conversion losses.

In summary, the system design and the adoption of MLPEs must be strategically considered based on detailed environmental and economic analyses. The findings confirm that while the performance gains are marginal in periods of low or uniform shading, the effective application of MLPEs can significantly enhance system energy yield under the partial shading conditions common in urban environments. However, the analysis also highlights a trade-off: despite the slight performance edge of microinverters, power optimizers offer greater design flexibility - such as selective deployment on only the most affected modules without altering the central inverter architecture - making them a competitive alternative from an investment and system renovation perspective. These findings highlight the critical role of using high-resolution models during the PV system design phase to accurately predict these performance trade-offs and optimize the final system configuration.

# 5. Discussion

## 5.1 Significance

This research marks a significant progression in the modeling of PV systems, particularly within complex urban contexts where partial shading frequently impairs performance. The hierarchical modeling framework developed has been rigorously validated, with its predictions of operational electrical characteristics (I-V-P) under varying partial shading conditions compared against field test data (Section 3). The results for both conventional PV modules and those equipped with MLPEs show remarkable accuracy, consistently achieving high $R^2$ scores (>0.90), which confirms the model's reliability. The true significance of this high-resolution approach, however, is most evident in its practical applications for both real-time forecasting and long-term system design, where traditional, coarser models exhibit critical shortcomings.

A primary significance lies in the framework's applicability to real-time power forecasting, a crucial requirement for effective distribution grid management. As demonstrated in Section 4.1, coarser resolution models not only significantly overestimate module MPPs (by up to 128%) and OPPs (by up to 149%) during suboptimal shading conditions (Figures 14, 15) but also frequently fail to identify when a module is completely bypassed. These per-module prediction errors were shown to propagate and aggregate at the string level, leading to OPP overestimations reaching as high as 163% (Figure 16 in Section 4.2). For distribution grid operators who depend on reliable short-term generation forecasts for grid balancing, voltage regulation, and stability, such inaccuracies are untenable. The proposed cell-level approach provides the necessary granularity to avoid these critical errors, enabling a more accurate and reliable prediction of PV system generation activities.

Furthermore, the framework's detailed approach is equally critical for accurate long-term energy performance modeling and optimal system design. The monthly energy analysis in Section 4.3 revealed that coarser models can overestimate the annual energy yield by miscalculating mismatch losses during seasons with frequent direct sunlight. For the heavily shaded string, the proposed cell-level approach effectively avoided overestimations of monthly energy yield that were reaching up to 54% with the coarser models (Figure 19). By accurately capturing these seasonal effects, the model serves as a reliable tool for assessing the cost-benefit of different system variants. It allows designers to quantitatively compare the annual energy gains of applying different MLPEs against conventional string or central inverter architectures (Figure 21), ensuring that system designs are optimized for maximum energy yield and financial return over their lifetime.

Finally, the hierarchical and modular nature of this research, culminating in a flexible Python-based tool package - PYWER (Figure 22), significantly enhances collaboration and innovation within the PV modeling community. By providing a validated, high-resolution foundation, the framework allows for tailored analyses specific to project needs - from optimizing large-scale solar farms to enhancing the efficiency of residential systems - thereby advancing not just technical capabilities in PV modeling but also empowering practical, evidence-based innovations in system design.

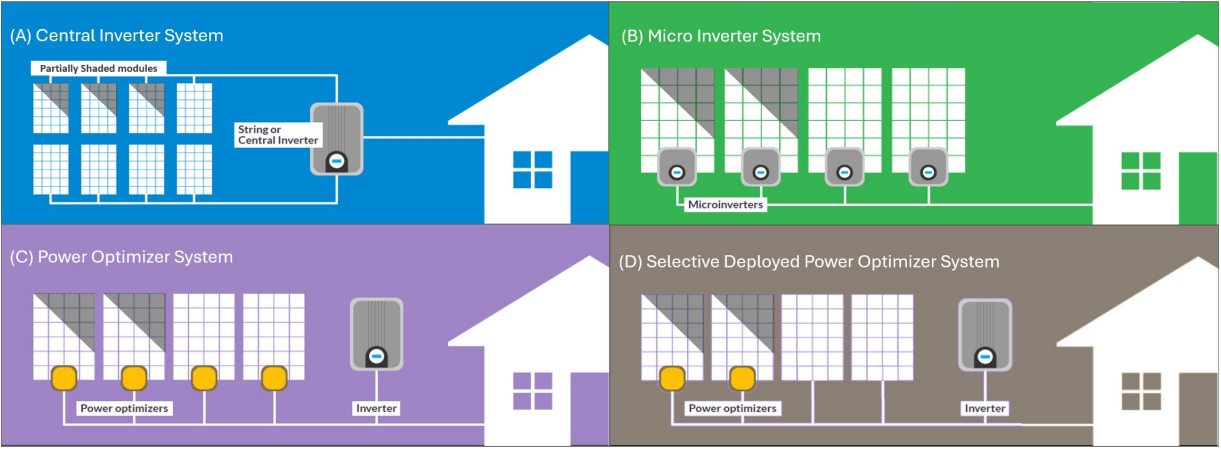

Figure 22. Supported PV system configurations for performance modeling in the developed PYWER tool package.

*5.2 Limitations and outlooks*

Despite its advantages, the complexity of high-resolution modeling necessitates increased computational resources, particularly for large-scale simulations. We observed that cell-level resolution is not always necessary for systems with minimal shading (Figure 19 in Section 4.3), suggesting that the "optimal resolution" for modeling may vary based on specific shading scenarios and system configurations. Future research should focus on developing adaptive resolution strategies that balance computational demands with modeling accuracy, providing guidelines for efficiently applying our methods across different settings.

Another limitation comes from the input data availability. The detailed irradiance and power models require multiple input data from the users; however, such information may not always be available during the early design stages, which in turn challenges the applicability of the method.

As the landscape of PV technology continues to evolve - through new materials, architectural innovations, and diverse installation strategies - updating our hierarchical models to incorporate these advancements will be critical. Moreover, expanding the model to integrate with other energy systems, such as storage solutions and grid interactions, introduces additional layers of complexity that must be addressed to enhance scalability and utility.

Currently, our MLPE extensions focus primarily on micro inverters and buck-type module power optimizers, covering a significant segment of the market yet omitting other vital technologies like boost optimizers. Future expansions should include a broader range of MLPE technologies, enhancing the tool's utility and encouraging community-driven enhancements within the PV sector.

Additionally, the irradiance modeling technique employed in this study does not fully account for external influences such as soiling, snow cover, and thermal gain of BIPV constructions, which potentially affect the accuracy of irradiance inputs. An important avenue for future development involves integrating these factors into our modeling workflow, refining the model's accuracy to better reflect real-world conditions and performance.

# 6. Conclusion

This study introduced and validated a pioneering high-resolution, hierarchical modeling framework capable of accurately assessing the performance of distributed PV systems in challenging urban environments. The framework's reliability was rigorously confirmed against field-test data, where its predictions of operational electrical characteristics for both conventional modules and those with MLPEs demonstrated remarkable accuracy, consistently achieving $R^2$ scores above 0.90.

A central finding of this research is the definitive, quantitative evidence of the limitations inherent in traditional, coarser-resolution PV models. The study demonstrates that the widely accepted trade-off between modeling simplicity and accuracy breaks down under the dynamic partial shading common in urban settings. Coarser-resolution models were shown to fundamentally misrepresent the system's operational state, leading to severe overestimations of both instantaneous operating power points (by up to 163%) and long-term monthly energy yields (by up to 54%). By capturing the non-linear effects of mismatch and the true bypassed state of modules, the proposed high-resolution approach provides a crucial correction to these inaccuracies.

The implications of this enhanced modeling accuracy are twofold. Firstly, it bridges the critical gap between optimistic design-phase predictions and complex operational reality. By enabling an evidence-based assessment of different system architectures and technologies, it empowers designers to reliably quantify the energy gains from MLPEs, optimize string configurations, and de-risk investments with more accurate financial forecasting. Secondly, the framework's ability to deliver high-fidelity, real-time power predictions provides a vital tool for advanced grid management applications, where reliable forecasting is essential for grid stability and the integration of distributed energy resources.

In conclusion, the high-resolution modeling approach presented in this study offers more than just an incremental improvement in accuracy; it provides a comprehensive and scalable tool for understanding and optimizing PV system performance. By faithfully representing the complex interplay between shading, hardware, and system architecture, this work supports more informed decision-making and empowers researchers and engineers to design more efficient, reliable, and economically viable PV systems in a variety of challenging installation contexts.

# Acknowledgments

The present work is sponsored by the scholarship from China Scholarship Council (CSC, 202107720038).

# Appendix A

*A.1 Supplement description of Fuentes cell temperature model*

As discussed in Section 2.2, the Fuentes model is predicated on the principle of energy balance, the core equation derived from energy balance is expressed as follow:

$$hc \cdot (T_c - T_a) + \varepsilon \cdot \sigma [(T_c^4 - T_s^4) + (T_c^4 - T_g^4)] - \alpha \cdot E_{eff} + m \cdot c \cdot \frac{dT_c}{dt} = 0 \qquad \text{A.1}$$

In the Equation A1, $hc$ represents the convective heat transfer coefficient [W/m²·K]. $T_c$, $T_a$, $T_s$, and $T_g$ denote the temperatures [K] of the cell, ambient air, sky, and ground respectively. The emissivity, $\varepsilon$, is a dimensionless measure of the module's efficiency in emitting thermal radiation compared to an ideal radiator, while $\sigma$ stands for the Stefan-Boltzmann constant, approximately 5.67 × 10⁻⁸ [W/m²·K⁴], governing the rate of heat loss through radiation. $\alpha$ represents the absorptivity of heat [-], and $E$ is the effective irradiance on the solar cell [W/m²]. Lastly, $m \cdot c$ signifies the thermal mass per cell unit area of the module [J/m²·K], influencing its thermal inertia, and $\frac{dT_c}{dt}$ is the rate of change of cell temperature over time [K/s].

A critical aspect of the Fuentes model is its use of the Installed Nominal Operating Cell Temperature (INOCT), which serves as a foundational temperature reference. The model then iteratively adjusts this baseline in response to the real-time environmental conditions, applying an energy balance equation to arrive at the operating cell temperature.

*A.2 Supplement description of solving two-diode equation*

In the present study, the solar cell modeling procedure begins by initializing PV module objects using module names or identifiers. Based on the defined module identifier, key modeling parameters will be queried from the California Energy Commission (CEC) solar equipment database [94], a widely recognized and authoritative source that contains standardized specification data for over 22000 commercially available modules. Table A.1 shows the retrieved parameter results of three example PV modules.

Table A.1. Reference parameters retrieved from CEC database.

| Retrieved parameters | Module names | | |
|---|---|---|---|
| | GermanSolar USA GSM6-60-300W | Centrosolar America EP72 335SW | First Solar Inc. FS-490A |
| Length $L$ [m] | 1.65 | 1.95 | 1.20 |
| Width $W$ [m] | 0.997 | 0.986 | 0.600 |
| Reference short circuit current $I_{SC}^{STC}$ [A] | 9.78 | 9.47 | 1.53 |
| Reference open circuit voltage $V_{OC}^{STC}$ [V] | 39.82 | 46.83 | 85.50 |
| Reference current at maximum power $I_{mp}^{STC}$ [A] | 9.33 | 8.84 | 1.36 |
| Reference voltage at maximum power $V_{mp}^{STC}$ [V] | 32.25 | 37.90 | 66.50 |
| Short circuit current temperature coefficient $\alpha_{Isc}$ [A/K] | 0.00255258 | 0.00521797 | 0.00091188 |
| Open circuit voltage temperature coefficient $\beta_{Voc}$ [V/K] | -0.135786 | -0.146578 | -0.22487 |
| Material band gap energy $E_g$ [eV] | 1.121 | 1.121 | 1.121 |
| Module type | Mono-crystalline | Multi-crystalline | Thin film |

Based on the retrieved module-specific parameters, we are then able to calculate and update the two-diode model parameters according to the irradiation and temperature conditions. We start with the determination of $V_t$ based on the cell temperature $T_c$ [K] as follow:

$$V_t = \frac{k \cdot T_c}{q} \quad \text{A.2}$$

Wherein $k = 1.38064852 \times 10^{-23}$ [J/K] is the Boltzmann constant, $q = 1.602176634 \times 10^{-19}$ [C] is the charge of an electron. Subsequently, the saturation currents for the first and second diodes at the cell operating temperature are given by:

$$I_{sat1} = I_{sat1}^{STC} \times \left(\frac{T_c}{T_{STC}}\right)^3 \times \exp\left(\frac{E_g \cdot q}{n_1 k} \cdot \left(\frac{1}{T_{STC}} - \frac{1}{T_c}\right)\right) \quad \text{A.3}$$

$$I_{sat2} = I_{sat2}^{STC} \times \left(\frac{T_c}{T_{STC}}\right)^3 \times \exp\left(\frac{E_g \cdot q}{n_2 \cdot k} \cdot \left(\frac{1}{T_{STC}} - \frac{1}{T_c}\right)\right) \quad \text{A.4}$$

Wherein $I_{sat1}^{STC}$ and $I_{sat2}^{STC}$ are the saturation currents of the two diodes under the standard test conditions with $T_{STC}$=298.15 [K]. Equation A.3 and A.4 can be further improved by taking the temperature variation influence on open circuit voltage into account [95], the improved equations express as:

$$I_{sat1} = \frac{I_{SC}^{STC} + \alpha_{Isc} \cdot (T_c - T_{STC})}{\exp\left[\frac{V_{OC}^{STC} + \beta_{Voc} \cdot (T_c - T_{STC})}{n_1 V_t}\right] - 1} \quad \text{A.5}$$

$$I_{sat2} = \frac{I_{SC}^{STC} + \alpha_{Isc} \cdot (T_c - T_{STC})}{\exp\left[\frac{V_{OC}^{STC} + \beta_{Voc} \cdot (T_c - T_{STC})}{n_2 V_t}\right] - 1} \quad \text{A.6}$$

The ideality factors and in Equation A.5 and A.6 can be determined based on the retrieved module type information (Table A.1). Specifically, we have compiled a datasheet of ideality factors from cutting-edge research [96–99], which has been curated and verified through extensive experiments and simulations. This allows our calculation workflow to automatically select the most appropriate, module type-specific values by referencing this collection (Table A.2).

Table A.2. Reference module type-specific ideality factors under various irradiance conditions, identified from literatures.

| Irradiance [W/m²] | Ideality factors | Module types | | |
| --- | --- | --- | --- | --- |
| | | Mono-crystalline | Multi-crystalline | Thin film |
| 200 | $n_1$ | 1.38 | 1.02 | 1.48 |
| | $n_2$ | 3.46 | 2.49 | 3.10 |
| 400 | $n_1$ | 1.38 | 1.02 | 1.44 |
| | $n_2$ | 2.30 | 2.69 | 3.68 |
| 600 | $n_1$ | 1.38 | 1.02 | 1.48 |
| | $n_2$ | 2.83 | 2.75 | 3.61 |
| 800 | $n_1$ | 1.38 | 1.03 | 1.46 |
| | $n_2$ | 3.15 | 2.62 | 3.31 |
| 1000 | $n_1$ | 1.37 | 1.03 | 1.48 |
| | $n_2$ | 2.11 | 2.35 | 3.72 |

Subsequently, during the calculation, the photogenerated current is adjusted for the ratio of effective irradiance $E_{eff}$ to the STC irradiance $E_{STC}$=1000 [W/m²] at each individual timestep:

$$I_{ph} = I_{SC}^{STC} \times (1 + \alpha_{Isc} \cdot (T_c - T_{STC})) \times \frac{E_{eff}}{E_{STC}} \qquad \text{A.7}$$

Furthermore, to estimate the series resistance ($R_s$) and shunt resistance ($R_{sh}$) parameters, we adapted the method by Ishaque et al. [100], employing an iterative process that leverages the comparison between estimated and actual maximum power points (MPP). Initially, $R_s$ is set to zero, and $R_{sh}$ is approximated using the voltage ($V_{mp}$) and current ($I_{mp}$) at MPP. With these initial conditions, we commence the iterative adjustment of $R_s$, increasing its value in each iteration, while $R_{sh}$ is recalculated accordingly. This iterative loop proceeds until the discrepancy between the estimated MPP and the true STC MPP, defined as the MPP error, falls below a predefined tolerance level at 2%. This approach ensures a precise calibration of $R_s$ and $R_{sh}$ based on actual MPP observations.

$$R_{sh\_initial} = \frac{V_{mp}^{STC}}{I_{SC}^{STC} - I_{mp}^{STC}} - \frac{V_{OC}^{STC} - V_{mp}^{STC}}{I_{mp}^{STC}} \qquad \text{A.8}$$

$$R_{sh} = \frac{V_{mp}^{STC} + I_{mp}^{STC} \cdot R_s}{I - I_{sat1}\left[\exp\left(\frac{V_{mp}^{STC} + I_{mp}^{STC} \cdot R_s}{n_1 \cdot V_t}\right) - 1\right] - I_{sat2}\left[\exp\left(\frac{V_{mp}^{STC} + I_{mp}^{STC} \cdot R_s}{n_2 \cdot V_t}\right) - 1\right] - I_{mp}^{STC}} \qquad \text{A.9}$$

Finally, the IV curve of a PV cell, based on the well-defined two-diode model, is calculated by applying the aforementioned equations across a range of voltage values. The procedure iteratively solves for the cell current $I$ at each voltage point $V$, taking into account the photogenerated current ($I_{ph}$), which varies with irradiance and temperature, as well as the voltage-dependent diode and resistive currents.

*A.3 Supplement description of modeling system IV curve from individual cells*

This appendix section provides the detailed methodology for deriving IV curves at each hierarchical level, starting from individual cells and extending to substrings, modules, and entire PV strings or systems. By following these stepwise procedures, one can capture partial shading and mismatch effects more accurately and better reflect real-world PV behavior.

*Cell substring IV curve.* The calculation of module level IV curves is based on the determined cell level IV curves. A general c-Si PV module comprises several sets of solar cells connected in series, so-called cell substrings. Under partial shading conditions, the cell with the lowest photocurrent ($I_{cell,\min}$) dictates the maximum current $I_{substring\_max}$ that can flow through an individual substring (Equation A.10). For $N$ solar cells with an individual substring, the substring level IV curve can be determined by iteratively aggregating the $V_i$ for each cell using the cell's IV curve for a given current $I$ (Equation A.11).

$$I_{substring\_max} = \min(I_{cell,1}, I_{cell,2}, \cdots, I_{cell,N}) \qquad \text{A.10}$$

$$V_{substring} = \sum_{i=1}^{N} V_i \qquad \text{A.11}$$

*Module IV curve.* Building upon the cell substring analysis, the next level involves calculating the IV curve for the entire PV module, which consists of multiple substrings connected in series configurations. Analogously, the approach to determine the module level IV curve is by aggregating the IV relations of all substrings. However, the calculation complexity arises at module level since each of the substrings has a bypass diode connection in parallel. The activated bypass diode will alter the current flow through the module under partial shading or fault conditions, affecting the overall module outputs. The conduction status of bypass diodes is determined by comparing the voltage across it $V_{diode}$ (Equation A.12) to the diode forward bias voltage $V_f$, which is typically around 0.7V for silicon diodes.

$$V_{diode} = V_{prev\_substring} - V_{next\_substring} \qquad \text{A.12}$$

If $V_{diode} \geq V_f$, the bypass diode is activated, and the corresponding substring does not contribute to the module current $I_{module}$. Afterwards, the $I_{module}$ is determined by the minimum current among the remaining active substrings that are not bypassed by activated diodes (Equation A.13). The module voltage can then be derived from sum of the voltage of active substrings and the forward voltage of the activated bypass diodes (Equation A.14).

$$I_{module} = \min(I_{active\_substring,\ 1},\ I_{active\_substring,\ 2},\ \cdots,\ I_{active\_substring,\ K}) \qquad \text{A.13}$$

$$V_{module} = \sum_{active} V_{substring} + N_{activated} \cdot V_f \qquad \text{A.14}$$

Based on the identified IV relations, the module level IV curve can be determined by iteratively increasing the current from zero (open circuit) towards the short circuit current, the changes of voltage across the module can be disclosed. It should be noted that initially all substrings would be contributing, but as the current increases, the voltage across some substrings might drop enough to activate their bypass diodes, removing those substrings from the current path and altering the module's overall IV characteristics.

*System IV curve.* The aggregation of modules into strings represents the IV curve modeling at system level. A module string is a series connection of several modules, aiming to match the system's voltage requirements. Understanding the transition from module IV curves to string IV curves is paramount for system design and integration. Analogous to the calculation of cell string IV curve, the current through the modules within the same string $I_{string}$ is consistent and limited by the weakest perform module (Equation A.15). The string voltage $V_{string}$ is the sum voltage across each module (Equation A.16).

$$I_{string} = \min(I_{module,1},\ I_{module,2},\ \cdots,\ I_{module,M}) \quad \text{A.15}$$

$$V_{string} = \sum_{i=1}^{M} V_{module,i} \quad \text{A.16}$$

For the larger-scale PV system that has multiple parallel-connected module strings in the configuration, its electrical characteristics can be determined by further scale up the IV curves of the module strings. As shown in Equation A.17 and A.18, the system output current $I_{system}$ is dictated by the cumulation of the string current, while the system voltage $V_{system}$ remains equal to that of an individual string.

$$I_{system} = \sum_{i}^{S} I_{string,i} \quad \text{A.17}$$

$$V_{system} = V_{string} \quad \text{A.18}$$

### A.4 Supplement description of PV power optimizer model

As mentioned in Section 2.4, power optimizers are typical MLPE devices designed to mitigate power losses in PV systems caused by module mismatch, partial shading, and other non-uniform operating conditions. This supplement section presents the theoretical details of the power optimizer model.

The model implements a dual-mode DC-DC converter that dynamically switches between operating modes based on the PV module's instantaneous conditions. This approach enables us to accurately simulate how power optimizers extract maximum power from each PV module independently, regardless of the operating conditions of other modules in the same string. The selection between these two modes is governed by the following decision rule:

$$\text{Mode} = \begin{cases} \text{Buck Converter,} & \text{if } I_{module} > I_{activate} \\ \text{Conductive,} & \text{if } I_{module} < I_{activate} \end{cases} \quad \text{A.19}$$

Where:

- $I_{module}$ represents the instantaneous current produced by the PV module;
- $I_{activate}$ is the current threshold that triggers the mode switch, typically set to the module's MPP current $I_{mpp}$.

This dual-mode architecture is crucial for maintaining high system efficiency across all operating conditions. During normal operation, the buck converter mode optimizes power extraction, while the conductive mode prevents unnecessary conversion losses during low-irradiance conditions. In the following sections, we will focus primarily on modeling the buck converter mode, as it represents the state of active power optimization. The conductive mode simply passes the module's electrical characteristics unchanged to the output, requiring no additional mathematical transformation.

### A.4.1 Impedance matching principle

The fundamental purpose of a power optimizer in buck converter mode is to perform impedance matching between the PV module and the rest of the system. This matching ensures maximum power transfer from the module to the load. The impedance matching is accomplished by adjusting the duty cycle of the buck converter. The theoretical optimal duty cycle (*D*) is derived from the relationship between input and output impedance within the converter:

$$D = \sqrt{\frac{Z_{out}}{Z_{in}}} = \sqrt{\frac{R_{out}}{R_{in}}} \qquad \text{A.20}$$

Where:

- $Z_{out}$ (or $R_{out}$) represents the output impedance of the converter, which corresponds to the load resistance at the MPP;
- $Z_{in}$ (or $R_{in}$) represents the input impedance, which is the dynamic resistance of the PV module at a given operating point.

This impedance matching principle forms the core of how power optimizers maintain each module at its individual maximum power point, regardless of the operating conditions of other modules in the system.

### A.4.2 Dynamic resistance calculation

For accurate impedance matching, we must calculate the dynamic resistance of the PV module at each point on its IV curve. The dynamic resistance is defined as the ratio of voltage to current at a specific operating point:

$$R_{in}(V, I) = \frac{V}{I} \qquad \text{A.21}$$

Similarly, we define the reference resistance at the MPP as:

$$R_{mpp} = \frac{V_{mpp}}{I_{mpp}} \qquad \text{A.22}$$

This reference resistance serves as a target for the impedance matching process. The optimizer adjusts its duty cycle to transform the actual module resistance to match this reference resistance, thereby ensuring operation at or near the MPP. This approach allows the model to account for the non-linear characteristics of PV modules and their response to varying environmental conditions.

### A.4.3 Duty cycle calculation with hardware constraints

Based on the impedance matching principle, we calculate the required duty cycle for individual point on the IV curve that would transform the actual module resistance to the desired resistance. To determine the appropriate calculation method, we distinguish between different operational regions of the module's IV curve:

- **Buck region** refers to the section of the IV curve from short-circuit current to the MPP, where voltage increases moderately as current decreases, where the optimizer needs to transform a lower input impedance up to match the MPP impedance.
- **Conductive region** refers to the section from MPP to open-circuit voltage, a duty cycle of $D=1$ represents this pass-through state, where the output voltage equals the input voltage.

$$D_{req} = \begin{cases} \sqrt{\dfrac{R_{mpp}}{R_{in}}}, & \text{if Buck region} \\ \sqrt{\dfrac{R_{mpp}}{R_{mpp}}} = 1, & \text{if Conductive region} \end{cases} \qquad \text{A.23}$$

However, real power optimizer hardware imposes physical constraints on the minimum and maximum duty cycle achievable. To accurately model these limitations, we apply boundary conditions to the calculated required duty cycle:

$$D_{actual} = \begin{cases} D_{\min}, & \text{if } D_{req} < D_{\min} \\ D_{\max}, & \text{if } D_{req} > D_{\max} \\ D_{req}, & \text{otherwise} \end{cases} \qquad \text{A.24}$$

Where:

- $D_{min}$ is the minimum duty cycle achievable by the hardware, typically around 0.1 (10%);
- $D_{max}$ is the maximum duty cycle achievable by the hardware, typically around 0.95 (95%).

These hardware constraints significantly impact optimizer performance at extreme operating points, such as heavily shaded conditions or substantial module mismatch. The constrained duty cycle represents what a physical power optimizer would actually implement, ensuring that our simulation results reflect the behavior of real-world devices.

### A.4.4 Voltage conversion in buck mode

In buck converter mode, the fundamental relationship between input and output voltage is governed by the duty cycle. For each operating point of the PV module, we calculate the optimizer output voltage as:

$$V_{out} = V_{in} \cdot D_{actual} \qquad \text{A.25}$$

Where:

- $V_{out}$ is the output voltage of the optimizer;
- $V_{in}$ is the input voltage from the PV module.

This relationship illustrates one of the key functions of the buck converter: stepping down the input voltage to a level that enables the module to operate at its maximum power point while maintaining compatibility with the rest of the system.

### A.4.5 Current conversion in buck mode

Within the buck mode region of the IV curve, the current is limited by either the maximum power point constraints or the hardware current limits:

$$I_{out} = \min\left(\frac{P_{mpp}}{V_{out}}, I_{\max}\right) \cdot \eta(V_{in}, I_{in}, D_{actual}) \qquad \text{A.26}$$

Where:

- $I_{max}$ is the maximum current the optimizer hardware can handle (typically 10-15A);
- $\eta(V_{in}, I_{in}, D_{actual})$ is the conversion efficiency at the specific operating point.

### A.4.6 Efficiency calculation model

The conversion efficiency $\eta$ of the buck converter significantly impacts the overall performance of the optimizer and the entire PV system. To accurately model this efficiency, we account for all major power loss mechanisms in the DC-DC converter circuit. This comprehensive approach ensures that our simulation provides realistic performance predictions across all operating conditions. Specifically, conversion efficiency is defined as:

$$\eta(V, I, D) = \frac{P_{out}}{P_{in}} = \frac{V_{out} \cdot I_{out}}{V_{in} \cdot I_{in}} = \frac{P_{in} - P_{loss}}{P_{in}} \qquad \text{A.27}$$

Where:

- $P_{out}$ and $P_{in}$ are the output and input powers;
- $P_{losses}$ represents the total power losses in the converter.

The total power loss is calculated as the sum of losses in individual components:

$$P_{losses} = P_{loss,switch} + P_{loss,inductor} + P_{loss,capacitor} + P_{loss,control} \qquad \text{A.28}$$

Where:

- $P_{loss,switch}, P_{loss,inductor}, P_{loss,capacitor}$ represents the losses from switch, inductor, and capacitor within the power optimizer circuit, respectively;
- $P_{loss,control}$ represents the constant power consumption by control circuitry of power optimizer.

This breakdown allows us to account for how different operating conditions affect each loss mechanism, providing a more accurate overall efficiency model. For instance, switching losses increase with frequency and voltage, while conduction losses increase with current. By modeling each component separately, we can accurately predict efficiency across the entire operating range of the optimizer.

# Appendix B

*B.1 Irradiance distribution of the system of interest in Section 4.3*

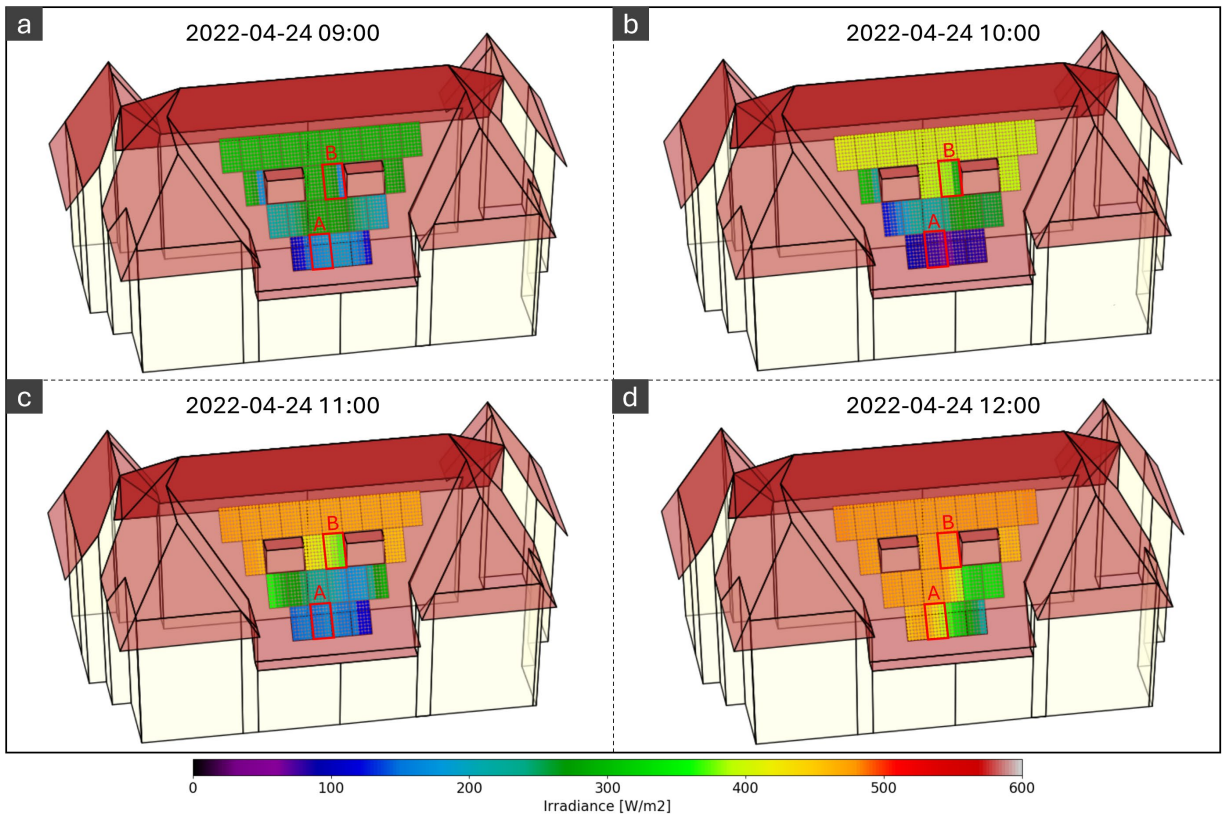

Figure B.1. Effective irradiance distribution that modeled by substring-level resolution at four different timestamps.

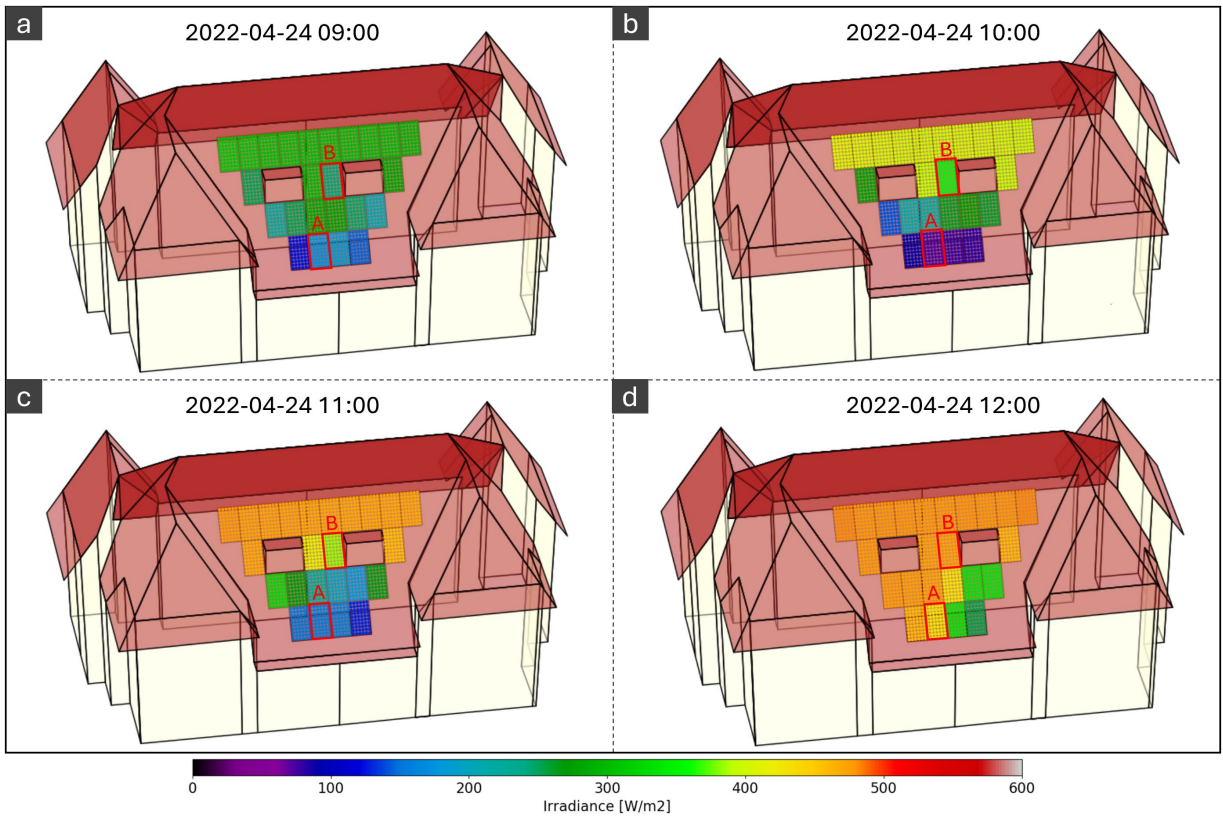

Figure B.2. Effective irradiance distribution that modeled by module-level resolution at four different timestamps.

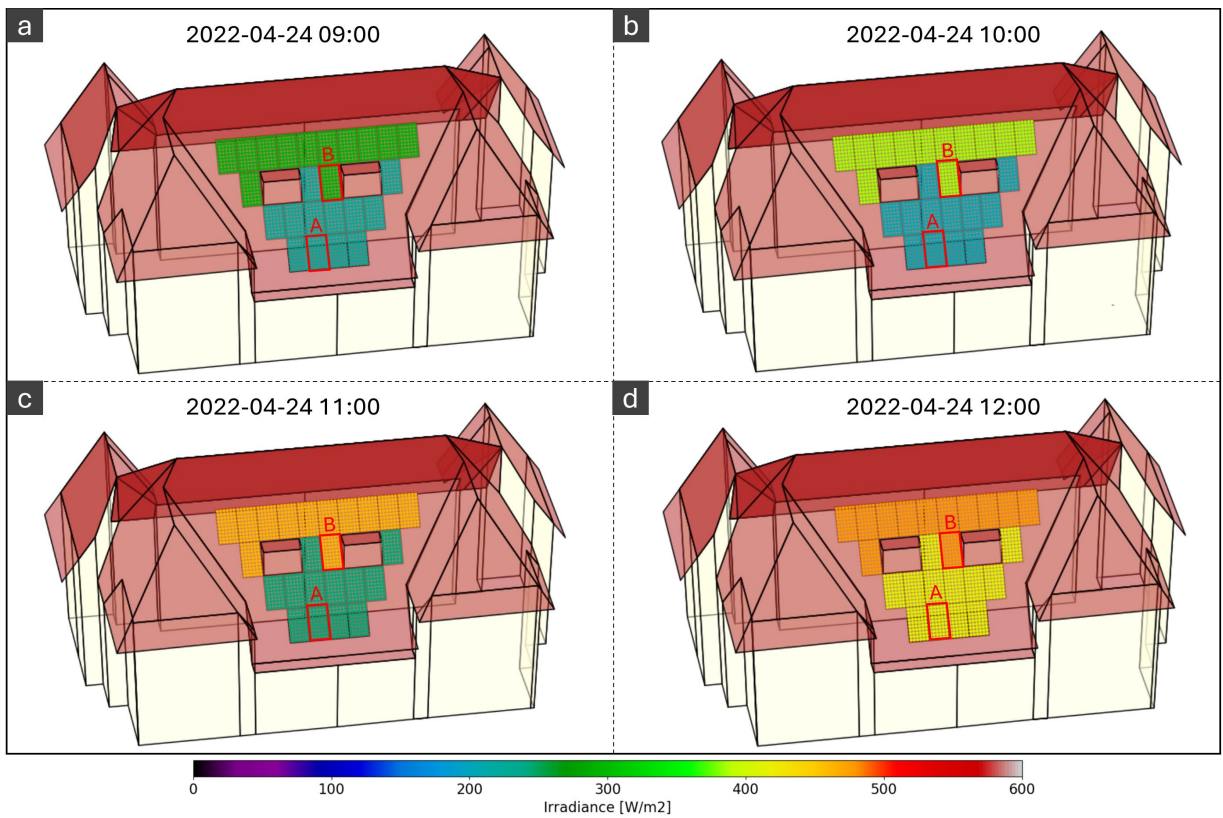

Figure B.3. Effective irradiance distribution that modeled by string-level resolution at four different timestamps.

## B.2 Supplement PV string IV curve modeling results

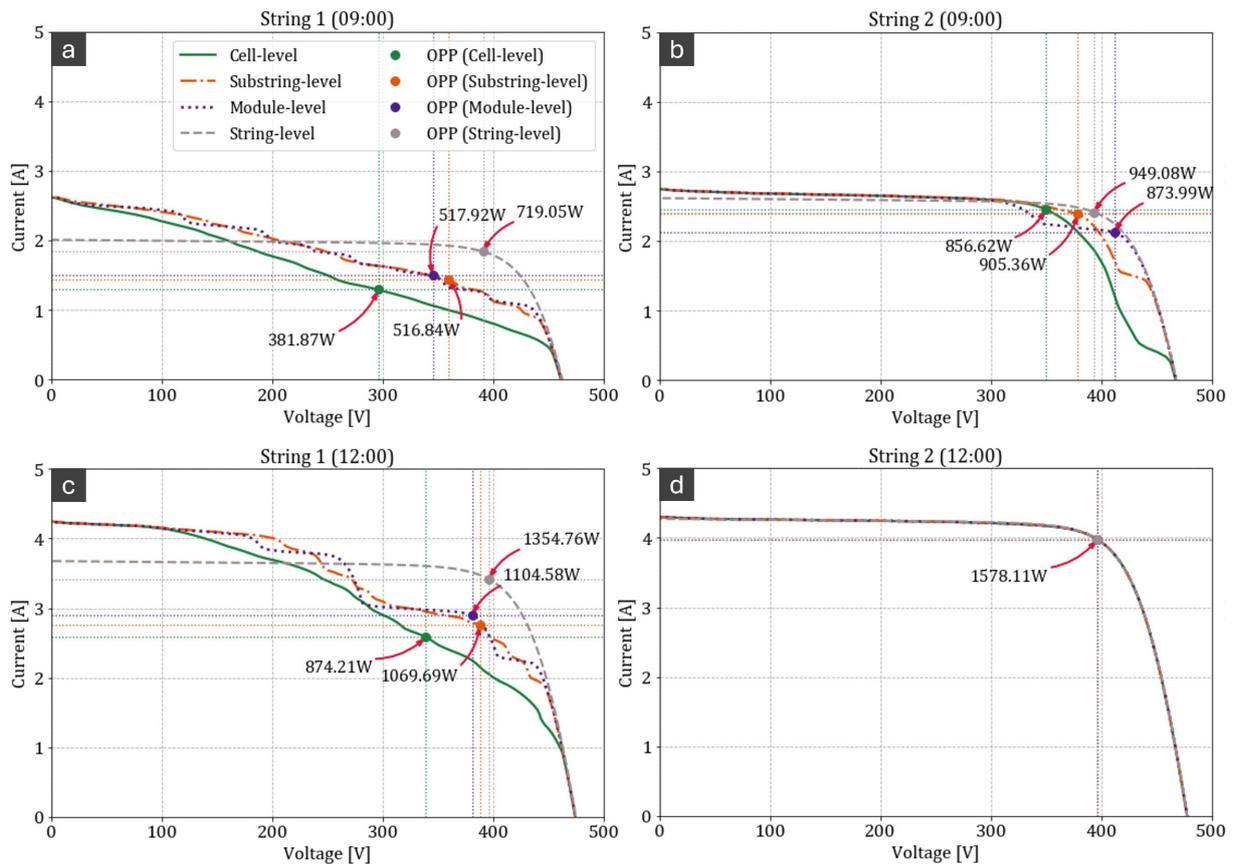

Figure B.4. String IV curves that modeled by different resolutions, where (a) string 1 at 9:00, (b) string 2 at 9:00, (c) string 1 at 12:00, and (d) string 2 at 12:00.

*B.3 Supplement PV string IV curves with or without optimizer equipped*

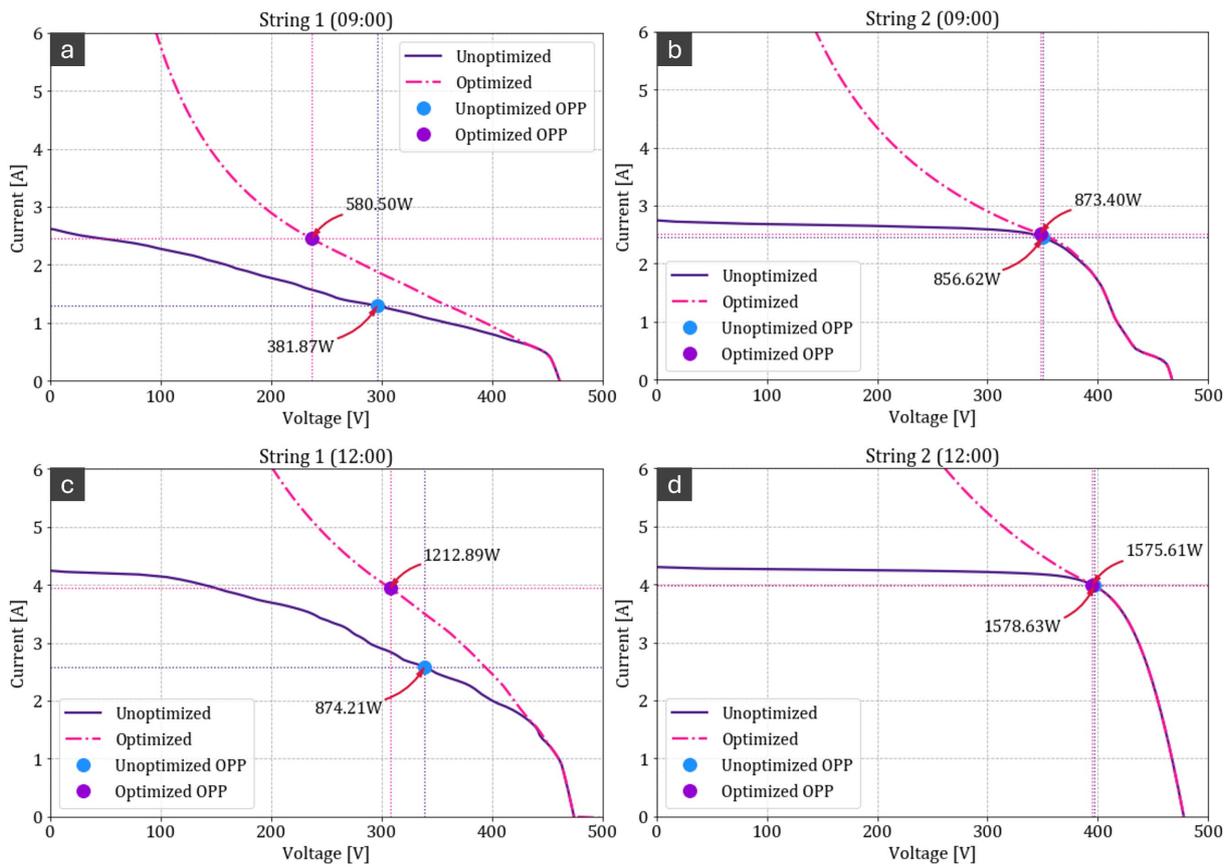

Figure B.5. String IV curves with and without power optimizer equipped that modeled by different resolutions, where (a) string 1 at 9:00, (b) string 2 at 9:00, (c) string 1 at 12:00, and (d) string 2 at 12:00.

# References

[1] IEA, Global Energy Review 2025, Paris, 2025. https://www.iea.org/reports/global-energy-review-2025.

[2] IEA, Electricity 2025, Paris, 2025. https://www.iea.org/reports/electricity-2025.

[3] M. Karmellos, G. Mavrotas, Multi-objective optimization and comparison framework for the design of Distributed Energy Systems, Energy Conversion and Management 180 (2019) 473–495. https://doi.org/10.1016/j.enconman.2018.10.083.

[4] IEA, Renewables 2024, Paris, 2024. https://www.iea.org/reports/renewables-2024.

[5] A.O. Ali, A.T. Elgohr, M.H. El-Mahdy, H.M. Zohir, A.Z. Emam, M.G. Mostafa, M. Al-Razgan, H.M. Kasem, M.S. Elhadidy, Advancements in photovoltaic technology: A comprehensive review of recent advances and future prospects, Energy Conversion and Management: X 26 (2025) 100952. https://doi.org/10.1016/j.ecmx.2025.100952.

[6] A. Calcabrini, R. Weegink, P. Manganiello, M. Zeman, O. Isabella, Simulation study of the electrical yield of various PV module topologies in partially shaded urban scenarios, Solar Energy 225 (2021) 726–733. https://doi.org/10.1016/j.solener.2021.07.061.

[7] P.R. Satpathy, R. Sharma, Power and mismatch losses mitigation by a fixed electrical reconfiguration technique for partially shaded photovoltaic arrays, Energy Conversion and Management 192 (2019) 52–70. https://doi.org/10.1016/j.enconman.2019.04.039.

[8] J. McCarty, C. Waibel, S.W. Leow, A. Schlueter, Towards a high resolution simulation framework for building integrated photovoltaics under partial shading in urban environments, Renewable Energy 236 (2024) 121442. https://doi.org/10.1016/j.renene.2024.121442.

[9] G. Raina, S. Sinha, A comprehensive assessment of electrical performance and mismatch losses in bifacial PV module under different front and rear side shading scenarios, Energy Conversion and Management 261 (2022) 115668. https://doi.org/10.1016/j.enconman.2022.115668.

[10] Algemeen Dagblad, Bomen illegaal gesnoeid en zelfs vergiftigd voor betere opbrengst van zonnepanelen | Binnenland | AD.nl, (2023). https://www.ad.nl/binnenland/bomen-illegaal-gesnoeid-en-zelfs-vergiftigd-voor-betere-opbrengst-van-zonnepanelen~a7b5d9e1/?cb=8050c15b-a96e-49be-a622-d706c95d6fcf&auth_rd=1 (accessed December 18, 2024).

[11] C. Wan, J. Zhao, Y. Song, Z. Xu, J. Lin, Z. Hu, Photovoltaic and solar power forecasting for smart grid energy management, CSEE Journal of Power and Energy Systems 1 (2015) 38–46. https://doi.org/10.17775/CSEEJPES.2015.00046.

[12] S. Sobri, S. Koohi-Kamali, N.Abd. Rahim, Solar photovoltaic generation forecasting methods: A review, Energy Conversion and Management 156 (2018) 459–497. https://doi.org/10.1016/j.enconman.2017.11.019.

[13] Vinod, R. Kumar, S.K. Singh, Solar photovoltaic modeling and simulation: As a renewable energy solution, Energy Reports 4 (2018) 701–712. https://doi.org/10.1016/j.egyr.2018.09.008.

[14] Y.A. Mahmoud, W. Xiao, H.H. Zeineldin, A Parameterization Approach for Enhancing PV Model Accuracy, IEEE Transactions on Industrial Electronics 60 (2013) 5708–5716. https://doi.org/10.1109/TIE.2012.2230606.

[15] C. Zhang, Y. Zhang, J. Su, T. Gu, M. Yang, Modeling and Prediction of PV Module Performance Under Different Operating Conditions Based on Power-Law I–V Model, IEEE Journal of Photovoltaics 10 (2020) 1816–1827. https://doi.org/10.1109/JPHOTOV.2020.3016607.

[16] Md.A. Islam, A. Merabet, R. Beguenane, H. Ibrahim, Modeling solar photovoltaic cell and simulated performance analysis of a 250W PV module, in: 2013 IEEE Electrical Power & Energy Conference, 2013: pp. 1–6. https://doi.org/10.1109/EPEC.2013.6802959.

[17] S. Chowdhury, G.A. Taylor, S.P. Chowdhury, A.K. Saha, Y.H. Song, Modelling, simulation and performance analysis of a PV array in an embedded environment, in: 2007 42nd International Universities Power Engineering Conference, 2007: pp. 781–785. https://doi.org/10.1109/UPEC.2007.4469048.

[18] A.S. Yadav, A.K. Maurya, V. Mukherjee, Performance investigation of ShapeDoKu variant for PV formations under realistic assumptions of shading situations, Optik 303 (2024) 171728. https://doi.org/10.1016/j.ijleo.2024.171728.

[19] E.A. Sarquis Filho, C.A.F. Fernandes, P.J. da Costa Branco, A complete framework for the simulation of photovoltaic arrays under mismatch conditions, Solar Energy 213 (2021) 13–26. https://doi.org/10.1016/j.solener.2020.10.055.

[20] S.R. Pendem, S. Mikkili, Modeling, simulation and performance analysis of solar PV array configurations (Series, Series–Parallel and Honey-Comb) to extract maximum power under Partial Shading Conditions, Energy Reports 4 (2018) 274–287. https://doi.org/10.1016/j.egyr.2018.03.003.

[21] Y. Li, X.M. Chen, B.Y. Zhao, Z.G. Zhao, R.Z. Wang, Development of a PV performance model for power output simulation at minutely resolution, Renewable Energy 111 (2017) 732–739. https://doi.org/10.1016/j.renene.2017.04.049.

[22] W. Sprenger, H.R. Wilson, T.E. Kuhn, Electricity yield simulation for the building-integrated photovoltaic system installed in the main building roof of the Fraunhofer Institute for Solar Energy Systems ISE, Solar Energy 135 (2016) 633–643. https://doi.org/10.1016/j.solener.2016.06.037.

[23] L. Walker, J. Hofer, A. Schlueter, High-resolution, parametric BIPV and electrical systems modeling and design, Applied Energy 238 (2019) 164–179. https://doi.org/10.1016/j.apenergy.2018.12.088.

[24] P. Manganiello, M. Baka, H. Goverde, T. Borgers, J. Govaerts, A. van der Heide, E. Voroshazi, F. Catthoor, A bottom-up energy simulation framework to accurately compare PV module topologies under non-uniform and dynamic operating conditions, in: 2017 IEEE 44th Photovoltaic Specialist Conference (PVSC), 2017: pp. 3343–3347. https://doi.org/10.1109/PVSC.2017.8366588.

[25] T. Ma, H. Yang, L. Lu, Solar photovoltaic system modeling and performance prediction, Renewable and Sustainable Energy Reviews 36 (2014) 304–315. https://doi.org/10.1016/j.rser.2014.04.057.

[26] B. Tian, R.C.G.M. Loonen, Á. Bognár, J.L.M. Hensen, Impacts of surface model generation approaches on raytracing-based solar potential estimation in urban areas, Renewable Energy 198 (2022) 804–824. https://doi.org/10.1016/j.renene.2022.08.095.

[27] S. Subramaniam, Daylighting simulations with radiance using matrix-based methods, Lawrence Berke-Ley National Laboratory (2017).

[28] N. Charles, M. Kabalan, P. Singh, Open source photovoltaic system performance modeling with python, in: 2015 IEEE Canada International Humanitarian Technology Conference (IHTC2015), 2015: pp. 1–4. https://doi.org/10.1109/IHTC.2015.7238046.

[29] K. Atluri, S.M. Hananya, B. Navothna, Performance of Rooftop Solar PV System with Crystalline Solar Cells, in: 2018 National Power Engineering Conference (NPEC), 2018: pp. 1–4. https://doi.org/10.1109/NPEC.2018.8476721.

[30] D. Riley, C. Hansen, J. Stein, M. Lave, J. Kallickal, B. Marion, F. Toor, A Performance Model for Bifacial PV Modules, in: 2017 IEEE 44th Photovoltaic Specialist Conference (PVSC), 2017: pp. 3348–3353. https://doi.org/10.1109/PVSC.2017.8366045.

[31] L.B. Bosman, S.B. Darling, Performance modeling and valuation of snow-covered PV systems: examination of a simplified approach to decrease forecasting error, Environ Sci Pollut Res 25 (2018) 15484–15491. https://doi.org/10.1007/s11356-018-1748-1.

[32] J. Polo, M.C. Alonso-García, J.P. Silva, M. Alonso-Abella, Modelling the performance of rooftop photovoltaic systems under urban Mediterranean outdoor conditions, Journal of Renewable and Sustainable Energy 8 (2016) 013502. https://doi.org/10.1063/1.4942856.

[33] J. Peng, D.C. Curcija, L. Lu, S.E. Selkowitz, H. Yang, R. Mitchell, Developing a method and simulation model for evaluating the overall energy performance of a ventilated semi-transparent photovoltaic double-skin facade, Progress in Photovoltaics: Research and Applications 24 (2016) 781–799. https://doi.org/10.1002/pip.2727.

[34] Y. Kurdi, B.J. Alkhatatbeh, S. Asadi, H. Jebelli, A decision-making design framework for the integration of PV systems in the urban energy planning process, Renewable Energy 197 (2022) 288–304. https://doi.org/10.1016/j.renene.2022.07.001.

[35] McNeil A, Using Clustering Techniques to Optimize Panel Grouping for Large PV Arrays with Non-uniform Orientation and Shading Obstructions, Proceedings of the 16th IBPSA Conference (2019).

[36] J. Kim, H. Lee, M. Choi, D. Kim, J. Yoon, Power performance assessment of PV blinds system considering self-shading effects, Solar Energy 262 (2023) 111834. https://doi.org/10.1016/j.solener.2023.111834.

[37] J. Polo, N. Martín-Chivelet, M. Alonso-Abella, C. Alonso-García, Photovoltaic generation on vertical façades in urban context from open satellite-derived solar resource data, Solar Energy 224 (2021) 1396–1405. https://doi.org/10.1016/j.solener.2021.07.011.

[38] Y. Zhou, M. Verkou, M. Zeman, H. Ziar, O. Isabella, A Comprehensive Workflow for High Resolution 3D Solar Photovoltaic Potential Mapping in Dense Urban Environment: A Case Study on Campus of Delft University of Technology, Solar RRL 6 (2022) 2100478. https://doi.org/10.1002/solr.202100478.

[39] PVPMC, Sandia PV Array Performance Model, (2024). https://pvpmc.sandia.gov/modeling-guide/2-dc-module-iv/point-value-models/sandia-pv-array-performance-model/ (accessed December 21, 2024).

[40] NREL, PVWatts Calculator, (2024). https://pvwatts.nrel.gov/ (accessed December 20, 2024).

[41] B. Lohani, S. Ghosh, Airborne LiDAR Technology: A Review of Data Collection and Processing Systems, Proc. Natl. Acad. Sci., India, Sect. A Phys. Sci. 87 (2017) 567–579. https://doi.org/10.1007/s40010-017-0435-9.

[42] OpenTopography, Data Catalog, (2024). https://portal.opentopography.org/dataCatalog?group=global (accessed December 23, 2024).

[43] K. Brecl, M. Topič, Apparent performance ratio of photovoltaic systems—A methodology for evaluation of photovoltaic systems across a region, Journal of Renewable and Sustainable Energy 8 (2016) 043501. https://doi.org/10.1063/1.4955088.

[44] T. Huld, A.M.G. Amillo, Estimating PV Module Performance over Large Geographical Regions: The Role of Irradiance, Air Temperature, Wind Speed and Solar Spectrum, Energies 8 (2015) 5159–5181. https://doi.org/10.3390/en8065159.

[45] K. Brecl, M. Topič, Photovoltaics (PV) System Energy Forecast on the Basis of the Local Weather Forecast: Problems, Uncertainties and Solutions, Energies 11 (2018) 1143. https://doi.org/10.3390/en11051143.

[46] A. Ameur, A. Berrada, K. Loudiyi, M. Aggour, Forecast modeling and performance assessment of solar PV systems, Journal of Cleaner Production 267 (2020) 122167. https://doi.org/10.1016/j.jclepro.2020.122167.

[47] K. Fath, J. Stengel, W. Sprenger, H.R. Wilson, F. Schultmann, T.E. Kuhn, A method for predicting the economic potential of (building-integrated) photovoltaics in urban areas based on hourly Radiance simulations, Solar Energy 116 (2015) 357–370. https://doi.org/10.1016/j.solener.2015.03.023.

[48] Y. An, T. Chen, L. Shi, C.K. Heng, J. Fan, Solar energy potential using GIS-based urban residential environmental data: A case study of Shenzhen, China, Sustainable Cities and Society 93 (2023) 104547. https://doi.org/10.1016/j.scs.2023.104547.

[49] L. Romero Rodríguez, E. Duminil, J. Sánchez Ramos, U. Eicker, Assessment of the photovoltaic potential at urban level based on 3D city models: A case study and new methodological approach, Solar Energy 146 (2017) 264–275. https://doi.org/10.1016/j.solener.2017.02.043.

[50] J.A. Jakubiec, C.F. Reinhart, A method for predicting city-wide electricity gains from photovoltaic panels based on LiDAR and GIS data combined with hourly Daysim simulations, Solar Energy 93 (2013) 127–143. https://doi.org/10.1016/j.solener.2013.03.022.

[51] M.C. Brito, S. Freitas, S. Guimarães, C. Catita, P. Redweik, The importance of facades for the solar PV potential of a Mediterranean city using LiDAR data, Renewable Energy 111 (2017) 85–94. https://doi.org/10.1016/j.renene.2017.03.085.

[52] Z. Li, Z. Zhang, K. Davey, Estimating Geographical PV Potential Using LiDAR Data for Buildings in Downtown San Francisco, Transactions in GIS 19 (2015) 930–963. https://doi.org/10.1111/tgis.12140.

[53] L. Kurdgelashvili, J. Li, C.-H. Shih, B. Attia, Estimating technical potential for rooftop photovoltaics in California, Arizona and New Jersey, Renewable Energy 95 (2016) 286–302. https://doi.org/10.1016/j.renene.2016.03.105.

[54] M. Cenky, J. Bendik, P. Janiga, I. Lazarenko, Urban-Scale Rooftop Photovoltaic Potential Estimation Using Open-Source Software and Public GIS Datasets, Smart Cities 7 (2024) 3962–3982. https://doi.org/10.3390/smartcities7060153.

[55] H. Jiang, X. Zhang, L. Yao, N. Lu, J. Qin, T. Liu, C. Zhou, High-resolution analysis of rooftop photovoltaic potential based on hourly generation simulations and load profiles, Applied Energy 348 (2023) 121553. https://doi.org/10.1016/j.apenergy.2023.121553.

[56] T. Zhong, Z. Zhang, M. Chen, K. Zhang, Z. Zhou, R. Zhu, Y. Wang, G. Lü, J. Yan, A city-scale estimation of rooftop solar photovoltaic potential based on deep learning, Applied Energy 298 (2021) 117132. https://doi.org/10.1016/j.apenergy.2021.117132.

[57] X. Song, Y. Huang, C. Zhao, Y. Liu, Y. Lu, Y. Chang, J. Yang, An Approach for Estimating Solar Photovoltaic Potential Based on Rooftop Retrieval from Remote Sensing Images, Energies 11 (2018) 3172. https://doi.org/10.3390/en11113172.

[58] L. Molin, S. Ericson, D. Lingfors, J. Munkhammar, J. Lindahl, B.S. AB, Validation of a PV generation model for simulation of wide area aggregated distributed PV power generation that takes individual systems location and orientation into account, in: 40th European Photovoltaic Solar Energy Conference and Exhibition, 2023.

[59] JRC, Photovoltaic Geographical Information System (PVGIS), (2024). https://re.jrc.ec.europa.eu/pvg_tools/en/ (accessed December 20, 2024).

[60] NREL, EnergyPlus 22.2.0, (2024). https://github.com/NREL/EnergyPlus/releases/tag/v22.2.0 (accessed December 20, 2024).

[61] O. Lindberg, J. Arnqvist, J. Munkhammar, D. Lingfors, Review on power-production modeling of hybrid wind and PV power parks, Journal of Renewable and Sustainable Energy 13 (2021) 042702. https://doi.org/10.1063/5.0056201.

[62] D. Lingfors, J. Widén, Development and validation of a wide-area model of hourly aggregate solar power generation, Energy 102 (2016) 559–566. https://doi.org/10.1016/j.energy.2016.02.085.

[63] G. Spagnuolo, S.F. KOURO RENAER, D. Vinnikov, Photovoltaic module and submodule level power electronics and control, IEEE Transactions on Industrial Electronics 66 (2019) 3856–3859.

[64] S. Islam, A. Woyte, R. Belmans, P. Heskes, P.M. Rooij, R. Hogedoorn, Cost effective second generation AC-modules: Development and testing aspects, Energy 31 (2006) 1897–1920. https://doi.org/10.1016/j.energy.2005.10.028.

[65] S. Sarwar, M.Y. Javed, M.H. Jaffery, M.S. Ashraf, M.T. Naveed, M.A. Hafeez, Modular Level Power Electronics (MLPE) Based Distributed PV System for Partial Shaded Conditions, Energies 15 (2022) 4797. https://doi.org/10.3390/en15134797.

[66] K. Sinapis, G. Litjens, M. van den Donker, W. Folkerts, W. van Sark, Outdoor characterization and comparison of string and MLPE under clear and partially shaded conditions, Energy Science & Engineering 3 (2015) 510–519. https://doi.org/10.1002/ese3.97.

[67] F. Famoso, R. Lanzafame, S. Maenza, P.F. Scandura, Performance Comparison between Micro-inverter and String-inverter Photovoltaic Systems, Energy Procedia 81 (2015) 526–539. https://doi.org/10.1016/j.egypro.2015.12.126.

[68] H. Lee, M. Choi, J. Kim, D. Kim, S. Bae, J. Yoon, Partial shading losses mitigation of a rooftop PV system with DC power optimizers based on operation and simulation-based evaluation, Solar Energy 265 (2023) 112053. https://doi.org/10.1016/j.solener.2023.112053.

[69] C. Allenspach, F. Carigiet, A. Bänziger, A. Schneider, F. Baumgartner, Power Conditioner Efficiencies and Annual Performance Analyses with Partially Shaded Photovoltaic Generators Using Indoor Measurements and Shading Simulations, Solar RRL 7 (2023) 2200596. https://doi.org/10.1002/solr.202200596.

[70] K. Sinapis, K. Tsatsakis, M. Dörenkämper, W.G.J.H.M. van Sark, Evaluation and Analysis of Selective Deployment of Power Optimizers for Residential PV Systems, Energies 14 (2021) 811. https://doi.org/10.3390/en14040811.

[71] B. Tian, R.C.G.M. Loonen, J.L.M. Hensen, Combining point cloud and surface methods for modeling partial shading impacts of trees on urban solar irradiance, Energy and Buildings 298 (2023) 113420. https://doi.org/10.1016/j.enbuild.2023.113420.

[72] N. Martin, J.M. Ruiz, Calculation of the PV modules angular losses under field conditions by means of an analytical model, Solar Energy Materials and Solar Cells 70 (2001) 25–38. https://doi.org/10.1016/S0927-0248(00)00408-6.

[73] M.M. Rahman, M. Hasanuzzaman, N.A. Rahim, Effects of various parameters on PV-module power and efficiency, Energy Conversion and Management 103 (2015) 348–358. https://doi.org/10.1016/j.enconman.2015.06.067.

[74] L. de O. Santos, P.C.M. de Carvalho, C. de O.C. Filho, Photovoltaic Cell Operating Temperature Models: A Review of Correlations and Parameters, IEEE Journal of Photovoltaics 12 (2022) 179–190. https://doi.org/10.1109/JPHOTOV.2021.3113156.

[75] M.K. Fuentes, A simplified thermal model for Flat-Plate photovoltaic arrays, Sandia National Labs., Albuquerque, NM (USA), 1987. https://www.osti.gov/biblio/6802914 (accessed March 9, 2024).

[76] A. Dolara, S. Leva, G. Manzolini, Comparison of different physical models for PV power output prediction, Solar Energy 119 (2015) 83–99. https://doi.org/10.1016/j.solener.2015.06.017.

[77] L. Deville, M. Theristis, B.H. King, T.L. Chambers, J.S. Stein, Open-source photovoltaic model pipeline validation against well-characterized system data, Progress in Photovoltaics: Research and Applications 1–13 (2013). https://doi.org/10.1002/pip.3763.

[78] Á. Bognár, PV in urban context: Modeling and simulation strategies for analyzing the performance of shaded PV systems, Phd Thesis 1 (Research TU/e / Graduation TU/e), Eindhoven University of Technology, 2021.

[79] S. Hubbard, PN Junctions and the Diode Equation, Photovoltaic Solar Energy: From Fundamentals to Applications (2016) 54–66.

[80] M.A. Green, Solar cells: operating principles, technology, and system applications, Englewood Cliffs (1982).

[81] K. Ishaque, Z. Salam, H. Taheri, Syafaruddin, Modeling and simulation of photovoltaic (PV) system during partial shading based on a two-diode model, Simulation Modelling Practice and Theory 19 (2011) 1613–1626. https://doi.org/10.1016/j.simpat.2011.04.005.

[82] L. Castañer, S. Silvestre, Modelling Photovoltaic Systems Using PSpice, John Wiley and Sons, 2002.

[83] E. Karatepe, M. Boztepe, M. Çolak, Development of a suitable model for characterizing photovoltaic arrays with shaded solar cells, Solar Energy 81 (2007) 977–992. https://doi.org/10.1016/j.solener.2006.12.001.

[84] R.W. Erickson, D. Maksimovic, Fundamentals of power electronics, Springer Science & Business Media, 2007.

[85] I. Houssamo, F. Locment, M. Sechilariu, Maximum power tracking for photovoltaic power system: Development and experimental comparison of two algorithms, Renewable Energy 35 (2010) 2381–2387. https://doi.org/10.1016/j.renene.2010.04.006.

[86] J. Dadkhah, M. Niroomand, Optimization Methods of MPPT Parameters for PV Systems: Review, Classification, and Comparison, Journal of Modern Power Systems and Clean Energy 9 (2021) 225–236. https://doi.org/10.35833/MPCE.2019.000379.

[87] N. Femia, G. Petrone, G. Spagnuolo, M. Vitelli, Optimization of perturb and observe maximum power point tracking method, IEEE Transactions on Power Electronics 20 (2005) 963–973. https://doi.org/10.1109/TPEL.2005.850975.

[88] S.B. Santra, D. Chatterjee, K. Kumar, M. Bertoluzzo, A. Sangwongwanich, F. Blaabjerg, Capacitor Selection Method in PV Interfaced Converter Suitable for Maximum Power Point Tracking, IEEE Journal of Emerging and Selected Topics in Power Electronics 9 (2021) 2136–2146. https://doi.org/10.1109/JESTPE.2020.2986858.

[89] Power Engineering International, Yokogawa's WT1800 precision power analyzer offers innovative measurement functions across a wide range of applications - Power Engineering International, (2020). https://www.powerengineeringint.com/news/yokogawas-wt1800-precision-power-analyzer-offers-innovative-measurement-functions-across-a-wide-range-of-applications/amp/ (accessed June 15, 2025).

[90] AHN, Actueel Hoogtebestand Nederland, (2022). https://www.ahn.nl/ (accessed February 18, 2022).

[91] KNMI, Koninklijk Nederlands Meteorologisch Instituut, (2023). https://www.knmi.nl/home (accessed March 14, 2023).

[92] Q. Wang, W. Yao, J. Fang, M. Xu, Dynamic Characteristics Analysis of Distributed PV Plants with Panel-level DC Optimizers Under Severe Partial Shading Conditions, in: 2022 7th Asia

Conference on Power and Electrical Engineering (ACPEE), 2022: pp. 1909–1915. https://doi.org/10.1109/ACPEE53904.2022.9784079.

[93] M. Lefèvre, A. Oumbe, P. Blanc, B. Espinar, B. Gschwind, Z. Qu, L. Wald, M. Schroedter-Homscheidt, C. Hoyer-Klick, A. Arola, A. Benedetti, J.W. Kaiser, J.-J. Morcrette, McClear: a new model estimating downwelling solar radiation at ground level in clear-sky conditions, Atmospheric Measurement Techniques 6 (2013) 2403–2418. https://doi.org/10.5194/amt-6-2403-2013.

[94] CEC, Solar Equipment List - California Energy Commission, (2025). https://solarequipment.energy.ca.gov/Home/PVModuleList (accessed June 24, 2025).

[95] M.G. Villalva, J.R. Gazoli, E.R. Filho, Comprehensive Approach to Modeling and Simulation of Photovoltaic Arrays, IEEE Transactions on Power Electronics 24 (2009) 1198–1208. https://doi.org/10.1109/TPEL.2009.2013862.

[96] X. Ma, S. Bader, B. Oelmann, On the performance of the two-diode model for photovoltaic cells under indoor artificial lighting, IEEE Access 9 (2020) 1350–1361.

[97] T. Ghizlane, O. Hassan, B. Abdelkader, B. Omar, Double diode ideality factor determination using the fixed-point method, International Journal of Advanced Computer Science and Applications 10 (2019).

[98] D.F. Alam, D.A. Yousri, M.B. Eteiba, Flower Pollination Algorithm based solar PV parameter estimation, Energy Conversion and Management 101 (2015) 410–422. https://doi.org/10.1016/j.enconman.2015.05.074.

[99] O. Hachana, K.E. Hemsas, G.M. Tina, C. Ventura, Comparison of different metaheuristic algorithms for parameter identification of photovoltaic cell/module, Journal of Renewable and Sustainable Energy 5 (2013) 053122. https://doi.org/10.1063/1.4822054.

[100] K. Ishaque, Z. Salam, H. Taheri, Simple, fast and accurate two-diode model for photovoltaic modules, Solar Energy Materials and Solar Cells 95 (2011) 586–594. https://doi.org/10.1016/j.solmat.2010.09.023.